\documentclass[prd,aps,showpacs,nofootinbib,tightenlines]{revtex4}  
\usepackage{mathrsfs}
\usepackage{amsmath}
\usepackage{amssymb}
\usepackage{epsfig}
\usepackage{graphicx}
\usepackage{booktabs}
\usepackage{multirow}
\usepackage{subfigure}
\usepackage{color}
\begin{document}
\newcommand{\psl}{ p \hspace{-1.8truemm}/ }
\newcommand{\nsl}{ n \hspace{-2.2truemm}/ }
\newcommand{\vsl}{ v \hspace{-2.2truemm}/ }
\newcommand{\epsl}{\epsilon \hspace{-1.8truemm}/\,  }


\title{Studies  of the resonance components in the $B_s$ decays into charmonia plus kaon pair}
\author{Zhou Rui$^1$}\email{jindui1127@126.com}
\author{Ya Li$^2$}\email{liyakelly@163.com}
\author{Hong Li$^1$}\email{lihong6608@163.com}
\affiliation{$^1$  College of Sciences, North China University of Science and Technology, Tangshan, Hebei 063210,   China}
\affiliation{$^2$ Department of Physics, College of Science, Nanjing Agricultural University, Nanjing, Jiangsu 210095, China}

\date{\today}
\begin{abstract}
In this work, the decays of  $B_s$ meson to a charmonium state
and a $K^+K^-$ pair are carefully investigated in the perturbative QCD approach.
Following the latest fit from the LHCb experiment,
we restrict ourselves to the case where the produced $K^+K^-$ pair interact in isospin zero
$S$, $P$, and $D$ wave resonances
in the kinematically allowed mass window.
Besides the dominant contributions of the $\phi(1020)$ resonance in the $P$-wave and $f_2'(1525)$ in the $D$-wave,
other resonant structures in the high mass region  as well as the $S$-wave components are also included.
The invariant mass spectra for most of the resonances  in the $B_s\rightarrow J/\psi K^+K^-$ decay are well reproduced.
The obtained three-body decay branching ratios  can reach the order of $10^{-4}$,
which seem to be accessible in the near future experiments.
The associated polarization fractions of those  vector-vector and  vector-tensor modes are also predicted,
which are compared with the existing data from the LHCb Collaboration.

\end{abstract}


\maketitle

\section{Introduction}
The three-body mode $B_s\rightarrow J/\psi K^+K^-$  is of particular interest in searches for intermediate states in the $B_s$ decay chain.
Since the LHCb Collaboration \cite{prd87072004} found no obvious structures in the $J/\psi K^+$ invariant mass distribution,
the $B_s\rightarrow J/\psi K^+K^-$  decay proceeds predominantly via $B_s\rightarrow J/\psi R$
with the  quasi-two-body intermediate state $R$ subsequently decaying into $K^+K^-$.
For the concerned $B_s$ decay,  the $K^+K^-$ system   arise from pure $s\bar s$ source, and thus these resonances are isoscalar.
Taking into account the conservation of $P$-parity and $C$-parity,
the produced resonances are limited to quantum numbers $J^{PC}=0^{++},1^{--},2^{++},...$ with isospin $I=0$.
Among them, the largest component comes from the $\phi(1020)$ in a $P$-wave configuration \cite{pdg2018}.
Several Collaborations \cite{prd87072004,prd546596,prd88114006} have presented a measurement
of the $B_s\rightarrow J/\psi \phi(1020)$ mode with $\phi(1020)$ decays to $K^+K^-$.
The current world averages of the absolute branching ratio $\mathcal{B}(B_s\rightarrow J/\psi \phi(1020))$ can reach the order of $10^{-3}$.
Another $P$-wave resonance $\phi(1680)$, whose contribution has more than 2 statistical standard deviation $(\sigma)$ significance,
 is  also included  by the LHCb Collaboration \cite{prd87072004} in its best fit model.
 Two well known scalar resonances, the $f_0(980)$ and  $f_0(1370)$,
are observed in the $K^+K^-$ mass spectrum by LHCb \cite{prd87072004},
which is the only data set available so far for the $S$-wave resonant structures.

Contributions from $D$-wave resonances are known to be non-negligible in this decay.
The first observation of the decay sequence $B_s\rightarrow J/\psi f_2'(1525), f_2'(1525)\rightarrow K^+K^-$,
 was recently reported by the LHCb Collaboration \cite{prl108151801},
and later confirmed by the D0 Collaboration \cite{prd86092011}.
Subsequently,  the LHCb Collaboration \cite{prd87072004} have determined the final state composition of the
 decay channel using a modified Dalitz plot analysis where the decay angular distributions are included.
The best fit model includes a nonresonant component and eight resonance states,
whose absolute branching ratios are measured
relative to that of the normalization decay mode $B^+\rightarrow J/\psi K^+$.
 In contrast to hadron collider experiments,
the Belle Collaboration \cite{prd88114006} normalize to the absolute number
of $B^0_s \bar{B}^0_s$ pairs produced and also present a measurement of the entire $B_s\rightarrow J/\psi  K^+K^-$
components including resonant and nonresonant decays.
More recently,  the LHCb Collaboration \cite{jhep080372017} improved their measurements, in which
 the fit fractions of six resonances including $\phi(1020), \phi (1680), f_2(1270), f_2'(1525), f_2(1750), f_2(1950)$
 together with  a $S$-wave structure in $B_s\rightarrow J/\psi  K^+K^-$ are determined.

Above measurements have caught  theoretical attention recently.
The three-body decay $B_s\rightarrow J/\psi  K^+K^-$ including  its dominant contributions of the resonances
$\phi(1020)$ and $f_2'(1525)$ have been studied \cite{prd89095026,prd95036013}  and the associated  branching
ratios  have been obtained based on the framework of the factorization approach.
Some recent analyses \cite{prd90094006,plb73770,prd90114004}
had  been carried out for  $B/B_s$ decays into $J/\psi$ and the scalar, vector, and
tensor resonances using chiral unitary theory,
for which these states  are shown to be generated from the meson-meson interaction.
In Refs \cite{prd79074024,jhep09074,jhep02009},  the $K^+K^-$ $S$-wave contribution in the $\phi(1020)$ resonance region
is estimated to be of the order $1-10\%$,   in agreement with previous measurements from
LHCb \cite{prd87072004,LHCb2012002}, CDF \cite{prl109171802}, and ATLAS \cite{jhep12072}.
The significant $S$-wave effects  may affect measurements of the $CP$  violating phase $\beta_s$ \cite{prd79074024,jhep09074,prd82076006}.

In this paper,
we will consider the three-body $B_s$  decays involving charmonia and kaon pair in the final state
under the quasi-two-body approximation in the framework of perturbative QCD approach (PQCD) \cite{prl744388,plb348597}.
The factorization formalism for the three-body decays can be  simplified to that for the two-body cases
with the introduction of two-kaon distribution amplitudes (DAs), which
absorb the strong interaction related to the production of the two kaon system.
For the detailed description of the three-body nonleptonic $B$ decays in this approach, one can refer to \cite{plb561258,prd70054006}.
The PQCD approach so far, has been successfully applied to the studies of the resonance contributions
to the three-body $B/B_s$ decays in several recent papers
\cite{prd91094024,plb76329,plb788468,epjc77199,prd97033006,prd98113003,181112738,prd95056008,prd96036014,prd98056019,prd96093011}.
As advanced before, the decays under study are dominated by a series of resonances
in $S$, $P$ and $D$ waves, 
while contributions from resonances with spin greater than two are not expected
since they are well beyond the available phase space. 
Each partial wave contribution is parametrized into the corresponding timelike
form factors involved in the two-kaon DAs.
For each partial wave form factor, we adopt the form as a linear
combination of those resonances with the same spin.
In the present paper, we take into account the following resonances
\footnote{ In the following, we also use the abbreviation $f_0$, $\phi$, and $f_2$
to denote the $S$, $P$, and $D$-wave resonances  for simplicity.}:  
$f_0(980)$, $f_0(1370)$, $\phi(1020)$, $\phi(1680)$, $f_2(1270)$, $f'_2(1525)$, $f_2(1750)$,  $f_2(1950)$,
two scalar, two vector, and four tensor resonances in the context of the data presented in Refs \cite{prd87072004,jhep080372017}.
All resonances are commonly described by Breit Wigner (BW) distributions, except for the $f_0(980)$ state,
 which is modelled by a Flatt\'{e} function \cite{plb63228}.

Our presentation is divided as follows.  In Sec. \ref{sec:framework},
  we present our model kinematics and describe the two-kaon DAs in different partial waves.
The calculated branching ratios and polarizations for each resonance in the  considered three-body decays
as well as the numerical discussions are presented in Sec.  \ref{sec:results},
and finally, conclusions are drawn in Sec \ref{sec:sum}.
The factorization formulas for the  decay amplitudes are collected in the Appendix.

\section{Kinematics and the two-kaon distribution amplitudes  }\label{sec:framework}
\begin{figure}[!htbh]
\begin{center}
\vspace{1.5cm} \centerline{\epsfxsize=7 cm \epsffile{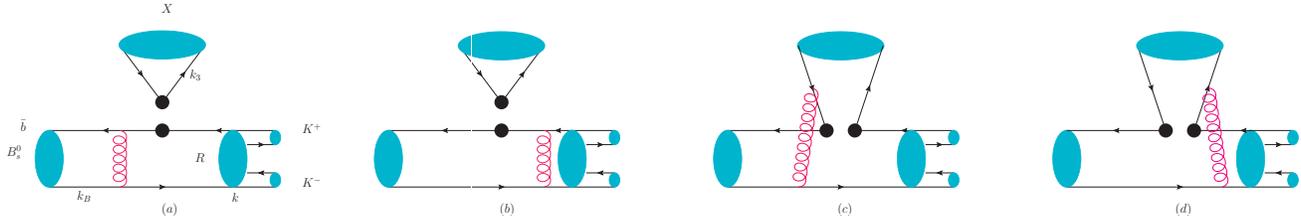}}
\caption{Feynman diagrams for the $B\rightarrow X R(\rightarrow K^+ K^-)$ decays with $X=J/\psi,\eta_c,\psi(2S),\eta_c(2S)$ at
the leading-order approximation. 
 (a) and (b) contributed to the factorizable  diagrams, while (c) and (d) contributed to the nonfactorizable ones.
The intermediate $R$ denotes a scalar, vector, or tensor resonance.}
 \label{fig:femy}
\end{center}
\end{figure}
Let us begin with the definition of the kinematic variables.
It is convenient to work in the light-cone coordinates for the four-momenta of the initial and final states.
The momentum  of the decaying $B_s$ meson in its rest frame is chosen as $ p_B=\frac{M}{\sqrt{2}}(1,1,\textbf{0}_{T})$
with the $B_s$ meson mass $M$.
The momenta of the decay products will be denoted as $p_1,p_2$ for the two kaons, and $p_3$ for the charmonia,
with the specific charge assignment according to
\begin{eqnarray}
B_s(p_B)\rightarrow X(p_3)K^+(p_1)K^-(p_2).
\end{eqnarray}
The momenta of three final states are defined   as
\begin{eqnarray}\label{eq:kin}
 p_1=(\zeta p^+, \eta (1-\zeta)p^+, \textbf{P}_{\text{T}} ),\quad 
 p_2=((1-\zeta)p^+, \eta \zeta p^+,-\textbf{P}_{\text{T}}),\quad
 p_3=\frac{M}{\sqrt{2}}(r^2,1-\eta,\textbf{0}_{\text{T}}),
\end{eqnarray}
where $\zeta=p_1^+/p^+$ with $p^+=M(1-r^2)/\sqrt{2}$ is the kaon momentum fraction.
The mass ratio $r=m/M$ with the charmonium mass $m$.
The kaon transverse momenta is expressed as $\textbf{P}_{\text{T}}=(\omega \sqrt{\zeta(1-\zeta)},0)$.
The factor $\eta=p^2/(M^2-m^2)$ is defined in terms of the invariant mass squared of the kaon pair $p^2=\omega^2$,
which satisfies the momentum conservation $p=p_1+p_2=p_B-p_3$.
The valence quark momenta labeled by $k_B$, $k_3$, and $k$, as indicated in Fig. \ref{fig:femy} (a),
are parametrized as
\begin{eqnarray}\label{eq:kkkk}
  k_B&=&(0,\frac{M}{\sqrt{2}}x_B,\textbf{k}_{\text{BT}}),\quad
  k_3=(\frac{M}{\sqrt{2}}r^2x_3,\frac{M}{\sqrt{2}}(1-\eta)x_3,\textbf{k}_{3\text{T}}),\quad
  k=(\frac{M}{\sqrt{2}}z(1-r^2),0,\textbf{k}_{\text{T}}),
\end{eqnarray}
in which $x_B$, $x_3$, $z$ denote the longitudinal momentum fractions,
and $k_{iT}$ represent the transverse momenta.
Since the light spectator quark momentum $k$ moves with the kaon pair in the plus direction, 
the minus component of its parton momentum should be very small,
thus it can be neglected in the hard kernel, and then integrated out
in the definition of the two-kaon distribution amplitudes.
We also dropped $k_B^+$ because it vanishes in the hard amplitudes.

Since the $B_s$ meson wave function and the charmonium distribution amplitudes have successfully described
various  hadronic two-body and three-body  charmonium  $B$ decays \cite{epjc77199,prd97033006,prd98113003,181112738,epjc77610},
we use the same ansatz as them.
For the sake of brevity, their explicit expressions are not shown here and can be found in Refs. \cite{epjc77610,prd90114030,epjc75293}.
Below, we briefly describe the two-kaon DAs in three partial waves and the associated form factors.
\subsection{$S$-wave two-kaon DAs}
The $S$-wave two-kaon DAs are introduced in analogy with the case of two-pion ones \cite{prd91094024,plb561258},
 which are organized into
\begin{eqnarray}\label{eq:fuliye2}
\Phi_{S}^{I=0}=\frac{1}{\sqrt{2N_c}}[\rlap{/}{p}\phi^0_S(z,\zeta,\omega^2)+
\omega\phi^s_S(z,\zeta,\omega^2)+\omega(\rlap{/}{n}\rlap{/}{v}-1)\phi^t_S(z,\zeta,\omega^2)],
\end{eqnarray}
with the null vectors  $n=(1,0,\textbf{0}_{\text{T}})$  and $v=(0,1,\textbf{0}_{\text{T}})$.
In what follows the subscripts $S$,  $P$, and $D$ always associate with the corresponding partial waves.
Above various twists DAs
have  similar  forms as the corresponding twists for a scalar meson
by replacing the scalar decay constant  with the scalar form factor \cite{plb730336},
we adopt their asymptotic models as shown below \cite{prd91094024,181112738}:
\begin{eqnarray}\label{eq:phi0st}
\phi^0_S(z,\zeta,\omega^2)&=&\frac{9}{\sqrt{2N_c}}F_S(\omega^2)a_1z(1-z)(1-2z), \nonumber\\
\phi^s_S(z,\zeta,\omega^2)&=&\frac{1}{2\sqrt{2N_c}}F_S(\omega^2),\nonumber\\
\phi^t_S(z,\zeta,\omega^2)&=&\frac{1}{2\sqrt{2N_c}}F_S(\omega^2)(1-2z),
\end{eqnarray}
with the isoscalar scalar form factor $F_S(\omega^2)$ and the Gegenbauer moment $a_1$.
Bearing in mind that   only odd moments contribute in case of  neutral scalar resonances
owing to charge conjugation invariance or conservation of vector current \cite{prd73014017}.
Therefore the first term in leading twist DAs come from $a_1$.
Since the coefficients in the Gegenbauer expansion of the dimeson DAs are poorly known,
we limit ourselves to leading term in the expansion.

For the scalar resonances, we include here only the components $f_0(980)$ and $f_0(1370)$,
which are well established  in the best fit model by the LHCb Collaboration \cite{prd87072004}.
For the former, we use a Flatt\'{e}  description, while the latter is modelled by BW  functions.
The scalar form factor $F_S(\omega^2)$ can be written as
\begin{eqnarray}
F_S(\omega^2)=[\frac{m^2_{f_0(980)}}{m^2_{f_0(980)}-\omega^2-i m_{f_0(980)}(g_{\pi\pi}\rho_{\pi\pi}+g_{KK}\rho_{KK}F_{KK}^2)}+c_{f_0(1370)}BW_{f_0(1370)}(\omega^2)](1+c_{f_0(1370)})^{-1}.
\end{eqnarray}
Hereafter,  $c_R$  refers to the  weight coefficient of the resonance $R$, to be determined by data.
Their values are given in the next section.
In what follows,  all resonances $R$ with different quantum numbers will be labeled by the single
letter $R$, without pointing to  its quantum numbers.
The two phase-space factors are $\rho_{\pi\pi}=2q_{\pi}/\omega$, $\rho_{KK}=2q_{K}/\omega$, where
 $q_{\pi(K)}$ is the pion (kaon) momentum in the dipion (dikaon) rest frame.
 The exponential factor $F_{KK}=e^{-\alpha q_K^2}$ with $\alpha=2.0\pm 0.25$ GeV$^{-2}$
  \cite{prd78074023,prd89092006} is introduced above the $KK$ threshold
and serves to reduce the  $\rho_{KK}$ factor as the invariant mass increases.
The constants $g_{\pi\pi}$ and $g_{KK}$ are the $f_0(980)$ couplings to $\pi\pi$ and $K\bar K$ final states respectively.
We use 
$g_{\pi\pi}=167$ MeV  and $g_{KK}/g_{\pi\pi}=3.47$ as determined by LHCb \cite{prd90012003}.
The BW amplitude in generic form is
\begin{eqnarray}
BW_{R}(\omega^2)=\frac{m_R^2}{m_R^2-\omega^2-im_R\Gamma(\omega^2)},
\end{eqnarray}
where $m_R$ is the resonance pole mass and $\Gamma(\omega^2)$ is its  energy-dependent width 
which may be parametrized in a form that ensures the correct behavior near threshold,
\begin{eqnarray}
\Gamma(\omega^2)=\Gamma_0(\frac{q_K}{q_{K0}})^{2L_R+1}\frac{m_R}{\omega}F^2_R.
\end{eqnarray}
Here $\Gamma_0$ and $q_{K0}$ are $\Gamma(\omega^2)$ and $q_{K}$, evaluated at the resonance pole mass, respectively.
$L_R$ is the orbital angular momentum in the $K^+K^-$ decay and is equal to the spin
of resonance $R$ because kaons have spin 0.  The $L_R=0,1,2,...$  correspond to the $S,P,D,...$ partial wave resonances.
The Blatt-Weisskopf barrier factors $F_R$ \cite{TNP} for scalar, vector and tensor states are
\begin{eqnarray}\label{eq:radii}
F_R=\left\{
\begin{aligned}
&1 \quad\quad\quad  &L_R=0, \\
&\frac{\sqrt{1+z_0}}{\sqrt{1+z}} \quad\quad\quad  &L_R=1, \\
&\frac{\sqrt{z_0^2+3z_0+9}}{\sqrt{z^2+3z+9}} \quad\quad\quad  &L_R=2, \\
\end{aligned}\right.
\end{eqnarray}
with $z=r^2q_K^2$ and $z_0$ represents the value of $z$ when $\omega=m_R$.
 The meson radius parameters $r$ are dependent on the momentum  of the decay particles in the parent rest frame.
Modifying the $r$ changes slightly our results, as discussed in the next section.
Hence, we set this parameter to be 1.5 $\text{GeV}^{-1}$ (corresponding to $0.3$ fm)
 for all the considered resonances, as is obtained in \cite{prd87072004}.

\subsection{$P$-wave two-kaon DAs}
In Ref.~\cite{prd98113003} we have constructed the $P$-wave DAs
including both longitudinal and transverse polarizations for the pion pair.
Naively, the $P$-wave two-kaon ones can be obtained by replacing the pion vector
form factors  by the corresponding kaon ones. The  explicit expressions read
\begin{eqnarray}\label{eq:pwavekk}
\Phi_{P}^{L}&=&\frac{1}{\sqrt{2N_c}}
[\rlap{/}{p}\phi^0_P(z,\zeta,\omega)+\omega \phi^s_P(z,\zeta,\omega)+
\frac{\rlap{/}{p}_1\rlap{/}{p}_2-\rlap{/}{p}_2\rlap{/}{p}_1}{\omega(2\zeta-1)}
\phi^t_P(z,\zeta,\omega)],\nonumber\\
\Phi_{P}^{T}&=&\frac{1}{\sqrt{2N_c}}
[\gamma_5\rlap{/}{\epsilon}_T\rlap{/}{p} \phi^T_P(z,\zeta,\omega)
+\omega \gamma_5\rlap{/}{\epsilon}_T \phi^a_P(z,\zeta,\omega)
+i\omega\frac{\epsilon^{\mu\nu\rho\sigma}\gamma_{\mu}
\epsilon_{T\nu}p_{\rho}n_{-\sigma}}{p\cdot n_-} \phi^v_P(z,\zeta,\omega)],
\end{eqnarray}
where the superscripts $L$ and $T$ on the left-hand side denote the longitudinal polarization and transverse  polarization, respectively.
Here $\epsilon^{\mu\nu\rho\sigma}$ is the totally antisymmetric unit Levi-Civita tensor with the convention $\epsilon^{0123}=1$.
The transverse polarization vector $\epsilon_{T}$ for the dikaon system has the same form as that of dipion \cite{prd98113003}.
The various twists DAs in Eq. (\ref{eq:pwavekk}) can be  expanded in terms of the Gegenbauer polynomials:
\begin{eqnarray}\label{eq:pwaveteist}
\phi^0_P(z,\zeta,\omega)&=&\frac{3F_P^{\parallel}(\omega^2)}
{\sqrt{2N_c}}z(1-z)[1+a^0_2C_2^{3/2}(1-2z)](2\zeta-1),\nonumber\\
\phi^s_P(z,\zeta,\omega)&=&\frac{3F_P^{\perp}(\omega^2)}
{2\sqrt{2N_c}}(1-2z)[1+a_2^s(1-10z+10z^2)](2\zeta-1),\nonumber\\
\phi^t_P(z,\zeta,\omega)&=&\frac{3F_P^{\perp}(\omega^2)}
{2\sqrt{2N_c}}(1-2z)^2[1+a^t_2C_2^{3/2}(1-2z)](2\zeta-1),\nonumber\\
\phi^T_P(z,\zeta,\omega)&=&\frac{3F_P^{\perp}(\omega^2)}
{\sqrt{2N_c}}z(1-z)[1+a^{T}_2C_2^{3/2}(1-2z)]\sqrt{\zeta(1-\zeta)},\nonumber\\
\phi^a_P(z,\zeta,\omega)&=&\frac{3F_P^{\parallel}(\omega^2)}
{4\sqrt{2N_c}}(1-2z)[1+a_2^a(10z^2-10z+1)]\sqrt{\zeta(1-\zeta)},\nonumber\\
\phi^v_P(z,\zeta,\omega)&=&\frac{F_P^{\parallel}(\omega^2)}
{2\sqrt{2N_c}}\{\frac{3}{4}[1+(1-2z)^2]+a_2^v[3(2z-1)^2-1]\}\sqrt{\zeta(1-\zeta)},
\end{eqnarray}
where the two $P$-wave form factors $F_P^{\parallel}$ and $F_P^{\perp}$ serve as the normalization of the two-kaon DAs.
They play a similar role with the vector and tensor decay constants in the definition of the vector meson DAs \cite{prd65014007}.
The Gegenbauer moments $a_2^i$ will be regarded as free parameters and determined in this work.

As mentioned in the Introduction, the form factor  $F^{\parallel}_P$ is given
by the coherence summation of the two vector resonances $\phi(1020)$ and $\phi(1680)$,
\begin{eqnarray}
F^{\parallel}_P(\omega^2)=[BW_{\phi(1020)}(\omega^2)+c_{\phi(1680)}BW_{\phi(1680)}(\omega^2)](1+c_{\phi(1680)})^{-1}.
\end{eqnarray}
 According to the argument in \cite{plb76329} [see Eq.(12)], motivated by the pole model,
each form factor is proportional to the decay constant associated with each resonance state. Therefore,
for the  $F^{\perp}_P$ , we assume it have the same phase as $F^{\parallel}_P$
and employ the approximate relation  $F_P^{\perp}/ F_P^{\parallel}\sim f^{T}_V/f_V$
 with $f^T_V$ ($f_V$)  being the tensor (vector) decay constant for the corresponding
 vector meson in the following calculations.
It is worth stressing that   the current data are still not sufficient to determine the two form factors separately.
 In principle, the longitudinal decay constant $f_V$ could be
extracted from the measurements, while the transverse one $f^{T}_V$ has to be calculated in the QCD
sum rule or Lattice QCD technique. It should be noted that the latter one is renormalization
scheme dependent and renormalization scale dependent, respectively. Then the approximate
relation $F_P^{\perp}/ F_P^{\parallel}$ could vary with the choices of decay constants at different energy scales.
In the numerical analysis, their values are chosen at the typical scale $\mu=1$ GeV, %
which enters the perturbative calculation in PQCD.

\subsection{$D$-wave two-kaon DAs}
Recalling that the tensor meson DAs are constructed in analogy with the vector ones by introducing a new
polarization vector $\epsilon_{\bullet}$ which is related to the polarization tensor
$\epsilon_{\mu\nu}(\lambda)$ with helicity $\lambda$ \cite{prd83014008}.
 Following a similar procedure,
we decompose the $D$-wave two-kaon DAs  associated with longitudinal
and transverse polarizations into
\begin{eqnarray}\label{eq:dwavekk}
\Phi_{D}^{L}&=&\sqrt{\frac{2}{3}}\frac{1}{\sqrt{2N_c}}
[\rlap{/}{p}\phi^0_D(z,\zeta,\omega)+\omega \phi^s_D(z,\zeta,\omega)+
\frac{\rlap{/}{p}_1\rlap{/}{p}_2-\rlap{/}{p}_2\rlap{/}{p}_1}{\omega(2\zeta-1)}
\phi^t_D(z,\zeta,\omega)],\nonumber\\
\Phi_{D}^{T}&=&\sqrt{\frac{1}{2}}\frac{1}{\sqrt{2N_c}}
[\gamma_5\rlap{/}{\epsilon}_{ T}\rlap{/}{p} \phi^T_D(z,\zeta,\omega)
+\omega \gamma_5\rlap{/}{\epsilon}_{ T} \phi^a_D(z,\zeta,\omega)
+i\omega\frac{\epsilon^{\mu\nu\rho\sigma}\gamma_{\mu}
\epsilon_{ T\nu}p_{\rho}n_{-\sigma}}{p\cdot n_-} \phi^v_D(z,\zeta,\omega)],
\end{eqnarray}
respectively, where the prefactor $\sqrt{\frac{2}{3}}$ ($\sqrt{\frac{1}{2}}$)
comes from the different definitions of the polarization vector between the vector and tensor mesons for the
longitudinal (transverse) polarization.
The leading twist DAs $\Phi_{D}^{0}$ and $\Phi_{D}^{T}$
have similar asymptotic forms as the corresponding ones for a tensor meson.
More precisely,
\begin{eqnarray}\label{eq:dwavetwist2}
\phi^0_D(z,\zeta,\omega)&=&\frac{9F^{\parallel}_D(\omega^2)}{\sqrt{2N_c}}z(1-z)a_1^0(2z-1)\mathcal{L}(\zeta),\nonumber\\
\phi^T_D(z,\zeta,\omega)&=&\frac{9F^{\perp}_D(\omega^2)}{\sqrt{2N_c}}z(1-z)a_1^T(2z-1)\mathcal{T}(\zeta).
\end{eqnarray}
Here,  we solely employ the first nonvanishing leading term in the expansion for previously mentioned reasons.
The moments $a_1^0$ and $a_1^T$ are regarded as  free parameters and determined in the next section.
Note that the $\zeta$ dependent terms $\mathcal{L}(\zeta)$ and $\mathcal{T}(\zeta)$ are different from those in Eq.~(\ref{eq:pwaveteist}).
We will derive their expressions later.
The kaon tensor form factor $F^{\parallel}_D(\omega^2)$ can be represented by
\begin{eqnarray}
F^{\parallel}_D(\omega^2)=  \sum_i c_iBW_i(\omega^2),
\end{eqnarray}
where the summation is performed over the intermediate tensor mesons: $f'_2(1525)$, $f_2(1270)$, $f_2(1750)$, $f_2(1950)$.
$c_i$ are the corresponding weight coefficients.
The  expressions for the twist-3 DAs  can be derived through the Wandzura-Wilczek relations as \cite{prd82054019,prd83034001}
\begin{eqnarray}\label{eq:dwavetwist3}
\phi^s_D(z,\zeta,\omega)&=&-\frac{9F^{\perp}_D(\omega^2)}{4\sqrt{2N_c}}a_1^0(1-6z+6z^2)\mathcal{L}(\zeta),\nonumber\\
\phi^t_D(z,\zeta,\omega)&=&\frac{9F^{\perp}_D(\omega^2)}{4\sqrt{2N_c}}(2z-1)a_1^0(1-6z+6z^2)\mathcal{L}(\zeta),\nonumber\\
\phi^a_D(z,\zeta,\omega)&=&\frac{3F^{\parallel}_D(\omega^2)}{2\sqrt{2N_c}}a_1^T(2z-1)^3\mathcal{T}(\zeta),\nonumber\\
\phi^v_D(z,\zeta,\omega)&=&-\frac{3F^{\parallel}_D(\omega^2)}{2\sqrt{2N_c}}a_1^T(1-6z+6z^2)\mathcal{T}(\zeta).
\end{eqnarray}

Next we derive  the $\zeta$ dependent terms for both the longitudinal and transverse polarization DAs.
The two decay constants $f_T$ and $f_T^{T}$ of a tensor meson are defined by sandwiching the 
corresponding local current operators between the vacuum and a tensor meson \cite{prd82054019,prd83014008}
\begin{eqnarray}\label{eq:decaycon}
\langle f_2(p,\lambda)|j_{\mu\nu}(0)|0\rangle&=&f_Tm_T^2\epsilon_{\mu\nu}^*(\lambda),\nonumber\\
\langle f_2(p,\lambda)|j_{\mu\nu\rho}(0)|0\rangle&=&-if^T_Tm_T[\epsilon_{\mu\rho}^*(\lambda)p_{\nu}-\epsilon_{\nu\rho}^*(\lambda)p_{\mu}],
\end{eqnarray}
where $p$ and $m_T$ are momentum and mass of  the tensor meson, respectively.
The two interpolating currents $j_{\mu\nu}(0)$ and $j_{\mu\nu\rho}(0)$ are defined in \cite{prd82054019,prd83014008}.
Let us begin with the local matrix element $\langle K^+(p_1)K^-(p_2)|j_{\mu\nu}(0)/j_{\mu\nu\rho}(0)|0\rangle$
 associated with  the $D$-wave  form factors.
Under the tensor-meson-dominant hypothesis \cite{prd82054019},
inserting the tensor intermediate  in above matrix element, we get
\begin{eqnarray}\label{eq:decaycon11}
\langle K^+(p_1)K^-(p_2)|j_{\mu\nu}(0)/j_{\mu\nu\rho}(0)|0\rangle\approx \sum_{\lambda}\langle K^+(p_1)K^-(p_2)|f_2(p,\lambda)\rangle
\frac{1}{\mathcal{D}_{f_2}}\langle f_2(p,\lambda)|j_{\mu\nu}/j_{\mu\nu\rho}|0\rangle,
\end{eqnarray}
with $\mathcal{D}_{f_2}$ the resonance propagator \cite{160503889}.
The coupling constant $g_{f_2KK}$ is defined by the matrix element $\langle K^+(p_1)K^-(p_2)|f_2(p,\lambda)\rangle=\frac{g_{f_2KK}}{m_T}\epsilon_{\mu\nu}(\lambda)q^{\mu}q^{\nu}$
with $q=p_1-p_2$ \cite{prd82054019}.
When applying the formula Eq.~(\ref{eq:decaycon}) and
the completeness relation  $\sum_{\lambda}\epsilon_{\mu\nu}(\lambda)\epsilon^*_{\rho\sigma}(\lambda)=
\frac{1}{2}M_{\mu\rho}M_{\nu\sigma}+\frac{1}{2}M_{\mu\sigma}M_{\nu\rho}-\frac{1}{3}M_{\mu\nu}M_{\rho\sigma}$
with $M_{\mu\nu}=g_{\mu\nu}-p_{\mu}p_{\nu}/m^2_T$,
Eq. (\ref{eq:decaycon11}) then leads to the following equations explicitly
\begin{eqnarray}
\label{eq:ll1}
\langle K^+(p_1)K^-(p_2)|j_{\mu\nu}|0\rangle &\approx & \frac{g_{f_2KK}f_Tm_T}
{\mathcal{D}_{f_2}}[q_{\mu}q_{\nu}-\frac{1}{3}p_{\mu}p_{\nu}  +\frac{1}{3}m_T^2g_{\mu\nu}  ], \\ \label{eq:ll2}
\langle K^+(p_1)K^-(p_2)|j_{\mu\nu\rho}|0\rangle &\approx &  -i\frac{g_{f_2KK}f_T^T}{\mathcal{D}_{f_2}}
[q_{\mu}p_{\nu}q_{\rho}-q_{\nu}p_{\mu}q_{\rho}  +\frac{1}{3}m_T^2(g_{\mu\rho} p_{\nu}-g_{\nu\rho} p_{\mu} )].
\end{eqnarray}
 Note that the last terms above are power suppressed  and can be omitted,
because in our power counting, a light hadron mass is counted as a low scale relative to the heavy quark mass.
Utilizing the approximation relation $q_{\mu}=(p_1-p_2)_{\mu}\approx (2\zeta-1)p_{\mu}$ \cite{prd98113003},
one get for Eq. (\ref{eq:ll1})
 \begin{eqnarray}
 q_{\mu}q_{\nu}-\frac{1}{3}p_{\mu}p_{\nu}=\frac{2}{3}(1-6\zeta+6\zeta^2)p_{\mu}p_{\nu},
 \end{eqnarray}
in which the coefficient $1-6\zeta+6\zeta^2$ is absorbed into the longitudinal  polarization  DAs,
giving rise to its $\zeta$ dependence.
The matrix element in Eq.~(\ref{eq:ll2}) for the choice $\mu,\nu,\rho=+,-,x$ is proportional to
 \begin{eqnarray}
(q_{\mu}p_{\nu}-q_{\nu}p_{\mu})q_{\rho}=2(2\zeta-1)\sqrt{\zeta(1-\zeta)}\omega^3,
 \end{eqnarray}
 where the kinematic variables in Eq.~(\ref{eq:kin}) are used.
Note that the contributions from other possible combination of the three Lorentz indexes  are either zero or power suppressed.
Then  the $\zeta$ dependent factors of the longitudinal and transverse polarization DAs  
can be written as
\begin{eqnarray}\label{eq:zeta}
\mathcal{L}(\zeta)=1-6\zeta+6\zeta^2, \quad \mathcal{T}(\zeta)=(2\zeta-1)\sqrt{\zeta(1-\zeta)},
\end{eqnarray}
respectively. Above expressions can also be checked from the  partial wave expansions of the helicity amplitudes.
As is well known, the helicity 0 component is expanded in terms of Legendre polynomials
$P_l(\text{cos} \theta)$, while the helicity $\pm1$ ones
proceeding in derivatives of the Legendre polynomials $P'_l(\text{cos} \theta)$ \cite{jhep02009}.
For $l=2$ $D$-wave amplitudes, the helicity angle $\theta$
is encoded into the Wigner-d functions, schematically \cite{pdg2018}:
 \begin{eqnarray}\label{eq:d00}
d_{00}^{l=2}(\theta)&=&P_2(\text{cos} \theta)=\frac{1}{2}(3\text{cos}^2 \theta-1), \nonumber\\
d_{\pm10}^{l=2}(\theta)&=&\mp \frac{\text{sin} \theta}{\sqrt{6}} P'_2(\text{cos} \theta)=\mp\sqrt{\frac{3}{2}}\text{sin} \theta \text{cos} \theta.
 \end{eqnarray}
Following a similar prescription in the dipion system \cite{npb905373,prd96051901},
the polar angle $\theta$ of the $K^+$ in the rest frame of the dikaon
is related to $\zeta$  through the relation
 \begin{eqnarray}\label{eq:cos}
1-2\zeta=\beta \text{cos} \theta, \quad \beta\equiv \sqrt{1-4m_K^2/p^2},
 \end{eqnarray}
 with $m_K$  the kaon  mass.
Neglecting the kaon mass and  employing  Eq. (\ref{eq:d00}), we also arrive at Eq. (\ref{eq:zeta}).

\subsection{The  differential branching ratio}
The double differential  branching ratio reads \cite{pdg2018}
 \begin{eqnarray}\label{eq:br}
\frac{d^2\mathcal{B}}{d \zeta d\omega}=\frac{\tau\omega|\vec{p}_1||\vec{p}_3|}{32\pi^3 M^3}|\mathcal{A}|^2,
 \end{eqnarray}
where the differential variable $d \text{cos} \theta$ is replaced by $d\zeta$ via Eq.~(\ref{eq:cos}) in the limit of massless.
The three-momenta of the kaon and  charmonium in the rest
reference frame of the $K\bar K$ system  are given by
\begin{eqnarray}
|\vec{p}_1|=\frac{\lambda^{1/2}(\omega^2,m_K^2,m_{K}^2)}{2\omega}, \quad
|\vec{p}_3|=\frac{\lambda^{1/2}(M^2,m^2,\omega^2)}{2\omega},
\end{eqnarray}
respectively, with  the standard K$\ddot{a}$ll$\acute{e}$n function $\lambda (a,b,c)= a^2+b^2+c^2-2(ab+ac+bc)$.
The complete amplitude $\mathcal{A}$ through resonance intermediate for the concerned decay is decomposed into
 \begin{eqnarray}\label{eq:amplitude}
\mathcal{A}=\mathcal{A}_S+\mathcal{A}_P+\mathcal{A}_D,
 \end{eqnarray}
where  $\mathcal{A}_S$, $\mathcal{A}_P$, and $\mathcal{A}_D$ denote the corresponding three partial wave decay amplitudes.
Since the $\zeta$-dependent terms  appear as an overall factor in each partial wave decay amplitudes,
integrating the double differential distribution of Eq. (\ref{eq:br}) over $\zeta$ gives for
the differential invariant mass branching ratio
 \begin{eqnarray}\label{eq:br2}
\frac{d\mathcal{B}}{d\omega}=\frac{\tau\omega|\vec{p}_1||\vec{p}_3|}{32\pi^3 M^3}
[|\mathcal{A}_S|^2+\frac{1}{3}|\mathcal{A}_{P}^{0}|^2+\frac{1}{5}|\mathcal{A}_{D}^{0}|^2+\sum_{j=\parallel,\perp}
(\frac{1}{6}|\mathcal{A}_{P}^{j}|^2+\frac{1}{30}|\mathcal{A}_{D}^{j}|^2)],
 \end{eqnarray}
where the factors $1/3,1/5,\cdots$ are extracted from the individual helicity amplitudes for the integral of $\zeta$.
The terms $\mathcal{A}^0$, $\mathcal{A}^{\parallel}$, and $\mathcal{A}^{\perp}$ represent the
longitudinal, parallel, and perpendicular polarization amplitudes in the transversity basis, respectively.
\footnote{ The last two terms do not appear for those modes involving spinless $\eta_c/\eta_c(2S)$ in the final state. }
Note that interference between different partial wave vanishes because the $\zeta$
functions in Eqs. (\ref{eq:phi0st}), (\ref{eq:pwaveteist}), and (\ref{eq:zeta}),
corresponding  to $S$, $P$, and $D$ partial waves, are orthogonal.
In the $J/\psi$ and $\psi(2S)$ cases,
the decay amplitudes $\mathcal{A}_S$ and $\mathcal{A}_P$ here can be straightforwardly obtained from the previous
publications \cite{prd91094024,prd98113003} by replacing  the two-pion form
factors and all pion masses and momenta with the respective kaon quantities.
For $\mathcal{A}_D$, its factorization formula can be related to $\mathcal{A}_P$
by making the following replacement,
 \begin{eqnarray}
\mathcal{A}_{D}^0=\sqrt{\frac{2}{3}}\mathcal{A}_{P}^0|_{\phi_{P}^{0,s,t}\rightarrow\phi_{D}^{0,s,t} },\quad
\mathcal{A}_{D}^{\parallel,\perp}=\sqrt{\frac{1}{2}}\mathcal{A}_{P}^{\parallel,\perp}|_{\phi_{P}^{T,v,a}\rightarrow\phi_{D}^{T,v,a} }.
 \end{eqnarray}
For the involved $\eta_c$ and $\eta_c(2S)$ modes, the partial wave decay amplitudes  are provided in Appendix.

\section{Numerical results}\label{sec:results}
\begin{table}
\caption{The relevant  resonance parameters in the $B_s\rightarrow X K^+K^-$ decays. }
\label{tab:resonant}
\begin{tabular}[t]{lccccc}
\hline\hline
Resonance & $J^{PC}$ & Resonance formalism & Mass (MeV) & Width (MeV) & Source  \\
\hline
$f_0(980)$ & $0^{++}$ & Flatt\'{e} &990&$\cdots$&PDG \cite{pdg2018}\\
$f_0(1370)$ & $0^{++}$ & BW &1475&113&LHCb \cite{prd86052006}\\
$\phi(1020)$ & $1^{--}$ & BW &1019&4.25&PDG \cite{pdg2018}\\
$\phi(1680)$ & $1^{--}$ & BW &1689&211 &Belle \cite{prd80031101}\\
$f_2(1270)$ & $2^{++}$ & BW &1276&187&PDG \cite{pdg2018}\\
$f'_2(1525)$ & $2^{++}$ & BW &1525&73&PDG \cite{pdg2018}\\
$f_2(1750)$ & $2^{++}$ & BW &1737&151&Belle \cite{epjc32323}\\
$f_2(1950)$ & $2^{++}$ & BW &1980&297&Belle \cite{epjc32323}\\
\hline\hline
\end{tabular}
\end{table}
We first summarize all parameter values required for numerical applications.
For the masses appearing in $B_s$ decays, we shall use the following values (in units of GeV) \cite{pdg2018}:
\begin{eqnarray}
M_{B_s}&=&5.367, \quad m_b=4.8, \quad m_c=1.275, \quad m_{K^{\pm}}=0.494, \nonumber\\
\quad m_{J/\psi}&=&3.097, \quad m_{\psi(2S)}=3.686, \quad m_{\eta_c}=2.984, \quad m_{\eta_c(2S)}=3.638.
\end{eqnarray}
The information on the   decay constants (in units of GeV),
the Wolfenstein parameters, together with the lifetime of $B_s$ mesons are adopted as \cite{prd82054019,prd98113003,epjc77610,prd90114030,epjc75293}
\begin{eqnarray}\label{eq:conf}
f_{B_s}&=&0.2272,  \quad  f_{J/\psi}=0.405, \quad  f_{\psi(2S)}=0.296, \quad  f_{\eta_c}=0.42, \quad  f_{\eta_c(2S)}=0.243,\quad f_{\phi(1020)}^T=0.186,\nonumber\\
f_{\phi(1020)}&=&0.215, \quad f_{f_2(1270)}=0.102, \quad
f_{f_2(1270)}^T=0.117,\quad f_{f'_2(1525)}=0.126, \quad f_{f'_2(1525)}^T=0.065, \nonumber\\
\lambda &=& 0.22537, \quad  A=0.814,  \quad \bar{\rho}=0.117, \quad \bar{\eta}=0.355,  \quad \tau_{B_s}=1.51 \text{ps}.
 \end{eqnarray}
The masses and widths of the BW resonances are listed in Table \ref{tab:resonant},
while the Flatt\'{e} parameters for the $f_0(980)$ have been given in  the previous section.

As mentioned before, the form factor ratio $r^T(R)=F^{\perp}/F^{\parallel}$ is approximately equal to the  ratio of two  decay constants $f^T/f$.
From the numbers in Eq.~(\ref{eq:conf}), we have
 \begin{eqnarray}\label{eq:rt1}
r^T(\phi(1020))=0.865, \quad r^T(f_2(1270))=1.15, \quad r^T(f'_2(1525))=0.52.
 \end{eqnarray}
Whereas for other high states, since their decay constants are not known yet,
we treat them as free parameters.

 Now, we collect all the phenomenologically motivated parameters,
 such as  weight coefficients $c_i$,  Gegenbauer moments $a_i$, and form factor ratios $r^T(R)$,
 in each partial wave. Their central values are fixed to be
\begin{eqnarray}\label{eq:rt2}
\text{S-wave}&:&  c_{f_0(1370)}=0.12e^{-i\frac{\pi}{2}},\quad a_1=0.8, \nonumber\\
\text{P-wave}&:&  c_{\phi(1680)}=0.6,\quad r^T(\phi(1680))=0.6,\quad a^0_2=a^{T}_2=-0.5,\quad
a^s_2=-0.7,\quad a^t_2=-0.3,\quad a^a_2=0.4,\quad a^v_2=-0.6,\nonumber\\
\text{D-wave}&:&  c_{f'_2(1525)}=1.2, \quad c_{f_2(1270)}=0.1e^{i\pi},  \quad c_{f_2(1750)}=0.4e^{i\pi},  \quad c_{f_2(1950)}=0.3, \nonumber\\
 && \quad r^T(f_2(1750))=0.3, \quad r^T(f_2(1950))=1.5, \quad a_1^0=0.4,\quad a_1^T=0.9.
 \end{eqnarray}

\begin{figure}[!htbh]
\begin{center}
\setlength{\abovecaptionskip}{0pt}
\centerline{
\hspace{4cm}\subfigure{\epsfxsize=13 cm \epsffile{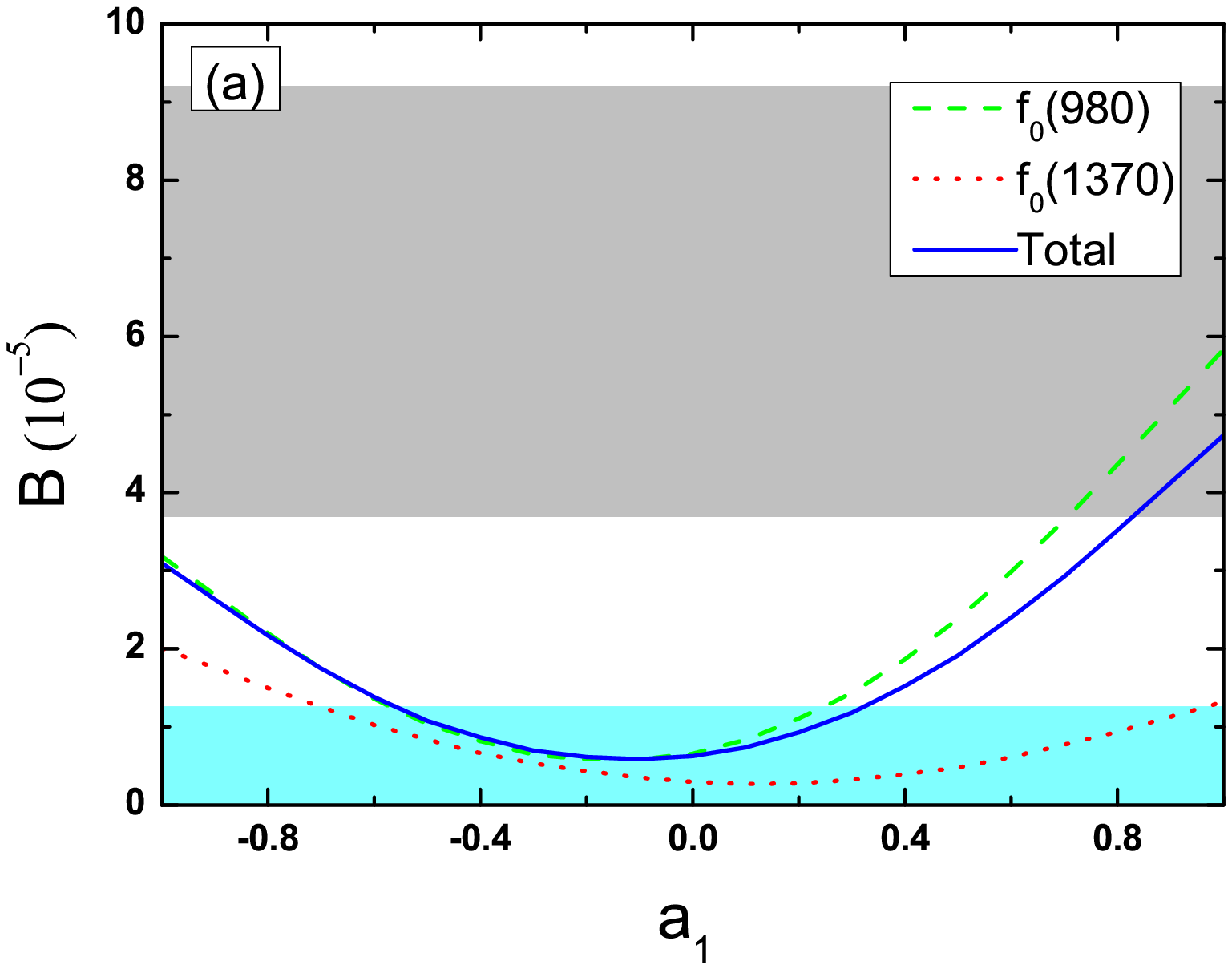} }
\hspace{-6cm}\subfigure{ \epsfxsize=13 cm \epsffile{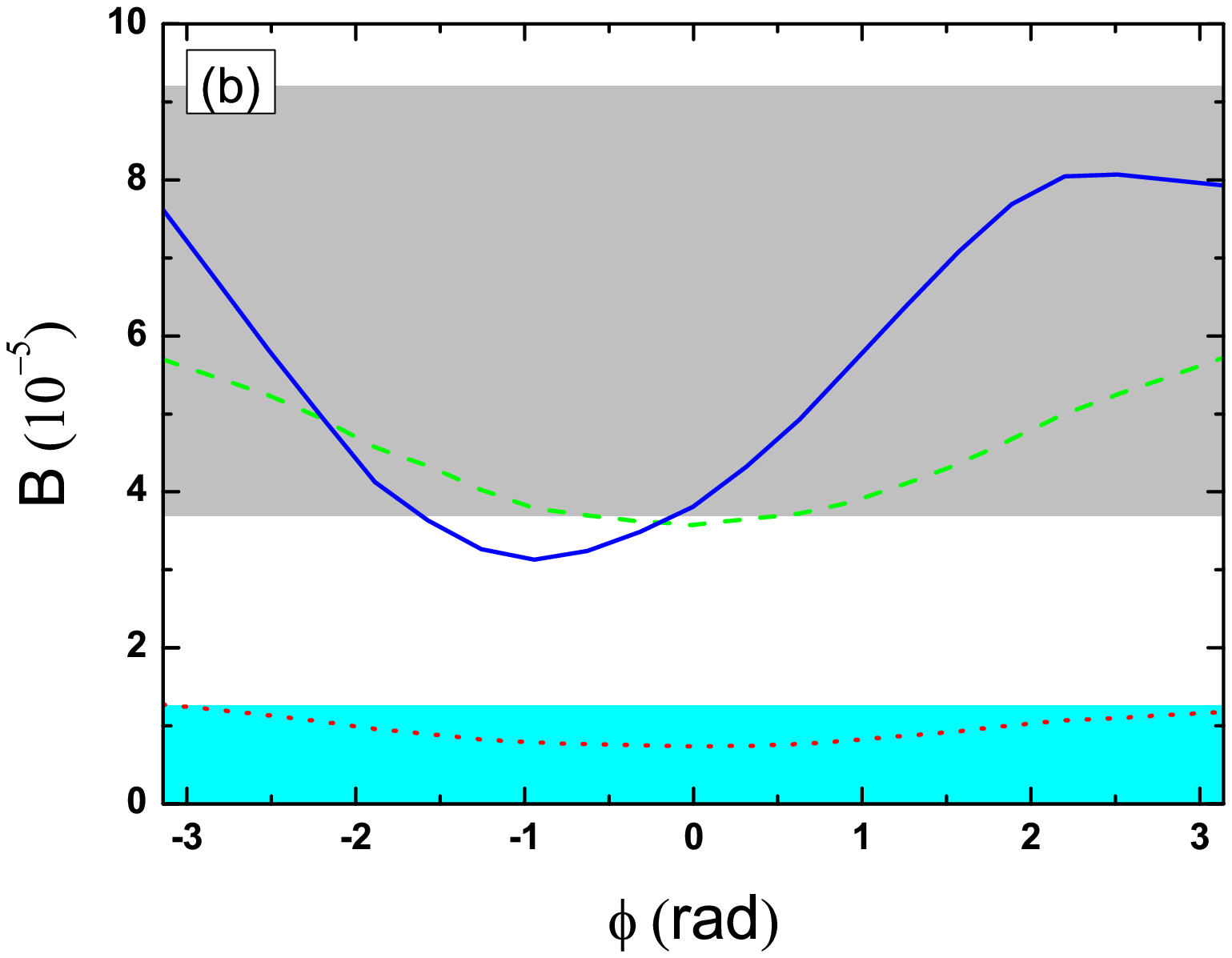}}}
\vspace{-3cm}\caption{
  The  branching ratios ($10^{-5}$) for the  $B_s\rightarrow J/\psi f_0(\rightarrow K^+K^-)$
  decays as a function of  (a) the Gegenbauer moment $a_1$ and (b) the phase $\phi$ of coefficient $c_{f_0(1370)}$
  with all other input fixed at the default  values in Eq.~(\ref{eq:rt2}).
 The  dashed green, dotted red,  and solid blue curves show the $f_0(980)$, $f_0(1370)$, and their combinatorial contributions, respectively.
 The gray and cyan shaded bands are corresponding to  the current experimental constraint of the
  $f_0(980)$ and $f_0(1370)$ modes from the LHCb \cite{prd87072004}, respectively.}
 \label{fig:a1phi}
\end{center}
\end{figure}

 When fitting  to the experimental data,
we assume that the concerned resonances with the same spin
share the same set of  the Gegenbauer moments.
For the $S$-wave sector,
the two experimental results of the $f_0(980)$ and $f_0(1370)$ components in Ref.~\cite{prd87072004}
are used to fit out $a_1$ and $c_{f_0(1370)}$.
In Fig.~\ref{fig:a1phi} (a) and (b), we show the dependence of
the branching ratios of the $f_0(980)$ and $f_0(1370)$ components as well as their combination
in the $B_s\rightarrow J/\psi K^+K^-$ decay on the Gegenbauer moment $a_1$ and the phase of $c_{f_0(1370)}$, respectively.
The module $|c_{f_0(1370)}|$ is chosen as 0.12
to maximize the overlap between the predicted curves and the experimental range.
Apparently, both the $f_0(980)$ and $f_0(1370)$ modes can meet the data as setting $a_1\sim 0.8$ in Fig.~\ref{fig:a1phi} (a).
This value is much larger than the corresponding $a_1=0.2$ that obtained in Ref.~\cite{prd91094024}.
The discrepancy is understandable with respect to the different nonperturbative dynamics of
 $f_0(980)$  decaying to the $KK$  and $\pi\pi$ pairs.
In Fig.~\ref{fig:a1phi} (b), it is  reflected that  the  constructive or destructive interference pattern
  between the two resonances vary with the phase.
Here its value is taken to be   $-\pi/2$ since the LHCb's data \cite{prd87072004} favor the destructive interference.
Of course, considering the sizeable experimental uncertainties,
these parameters are difficult to be restricted  precisely at this moment.
 As a case study with rough estimation, the related treatment about these
parameters in this work is just a try. A convincible research should be performed
through a global fit to more rich measurements in the future.

For the $P$-wave  ones, we first use the experimental branching ratios of three decay channels
$B_s\rightarrow J/\psi \phi(1020)(\rightarrow K^+K^-)$ (longitudinal) \cite{jhep080372017},
$B_s\rightarrow \psi(2S) \phi(1020)(\rightarrow K^+K^-)$ (longitudinal) \cite{pdg2018}, and
$B_s\rightarrow \eta_c  \phi(1020)(\rightarrow K^+K^-)$ \cite{jhep070212017}
to fit the three longitudinal Gegenbauer moments  $a_2^0$,  $a_2^s$, and $a_2^t$,
then the three transverse ones can be constrained  by the  transverse polarization fractions of the former two modes.
Finally, according to the fit fraction and polarizations of the $\phi(1680)$ component in the
 $B_s\rightarrow J/\psi K^+K^-$ decay from the LHCb \cite{jhep080372017},
 one can determine the values of $r^T(\phi(1680))$ and  $c_{\phi(1680)}$.

Since the $f'_2(1525)$ component in the $J/\psi$ mode is well measured with a relatively high accuracy
 comparing with other $D$-wave resonances by the LHCb Collaboration \cite{jhep080372017},
 we can exactly determine its weight coefficient $c_{f_2'(1525)}$ and two Gegenbauer moments $a_1^0$ and $a_1^T$
 based on its fit fraction and three polarizations. Following a similar procedure as above,
we can determine $r^T(R)$ and the module of $c_R$  for other tensor resonances.
As pointed out in \cite{epjc3941}, the form factor $F(s)$ with the time-like momentum transfer squared $s>4m_K^2$
could be analytically continued to the space-like region $s<0$.
It has been known that a form factor is normalized to unity at $s=0$,
because a soft probe cannot reveal the structure of a bound state.
Therefore, we postulate that the kaon form factors should be constrained by such normalization condition.
According to our  fitted  modules of the $D$-wave weight coefficients in Eq.~(\ref{eq:rt2}),
the phases of $c_{f_2(1270)}$ and $c_{f_2(1750)}$ are set to $\pi$ to ensure the normalization of the form factor $F^{\parallel}_D(0)=1$.
Strictly speaking, the phases of the various coefficient $c_i$ in Eq.~(\ref{eq:rt2})
should be determined from the interference fit fractions.
However, the current available data are not yet sufficiently precise to extract them.
Furthermore,
 the $f'_2(1525)$  dominates over the $D$-wave contributions as shown below,
the relative phases among these $c$ parameters have little effect on the total $D$-wave decay branching ratios,
and our choice of these phases does not affect the magnitude estimation of either the individual resonance or the total contribution.

\begin{table}
\caption{ Branching ratios  of $S$-wave resonance contributions to  the $B_s\rightarrow (J/\psi,\psi(2S),\eta_c,\eta_c(2S))  K^+K^-$ decays.
 Theoretical errors correspond to the uncertainties of Gegenbauer moments and hard scales, respectively.}
\label{tab:brs}
\begin{tabular}[t]{lccc}
\hline\hline
Modes & $\mathcal{B}(R=f_0(980))$& $\mathcal{B}(R=f_0(1370))$ & \text{S-wave}\footnotemark[1]\\ \hline
$J/\psi  K^+K^- $  & $(4.3_{-1.3-0.3}^{+1.3+0.7})\times 10^{-5}$
& $(9.4^{+3.1+1.5}_{-3.4-0.4})\times 10^{-6}$  & $(3.4_{-0.9-0.4}^{+1.0+0.5})\times 10^{-5}$  \\
Data \cite{prd87072004} & $(3.7\sim 9.2)\times 10^{-5}$ \footnotemark[2]
&$(9.2^{+3.3}_{-9.2})\times 10^{-6}$ \footnotemark[3] & $\cdots$ \\
$\eta_c  K^+K^- $  & $(4.6_{-1.1-0.5}^{+0.8+0.8})\times 10^{-5}$ & $(1.1^{+0.2+0.2}_{-0.3-0.1})\times 10^{-5}$
& $(3.5_{-0.7-0.3}^{+0.8+0.6})\times 10^{-5}$\\
$\psi(2S)  K^+K^- $  & $(8.0_{-2.6-1.2}^{+2.2+1.8})\times 10^{-6}$ & $(8.3^{+3.0+1.9}_{-3.0-0.8})\times 10^{-7}$
& $(6.2_{-1.7-0.8}^{+1.7+1.4})\times 10^{-6}$\\
$\eta_c(2S)  K^+K^-$  & $(9.3_{-1.9-1.1}^{+1.7+1.9})\times 10^{-6}$
& $(1.4^{+0.3+0.3}_{-0.3-0.1})\times 10^{-6}$& $(7.2^{+1.5+1.4}_{-1.3-0.9})\times 10^{-6}$ \\
\hline\hline
\end{tabular}
\footnotetext[1]{We are not including contributions from the nonresonant $S$-wave.}
\footnotetext[2]{We quote  the range of measurement since the fit fraction of $f_0(980)$ is strongly parametrization dependent. }
\footnotetext[3]{The fit fraction statistical and systematic are  added in quadrature.}
\end{table}
\begin{table}
\caption{  Branching ratios  of $P$-wave resonance contributions to  the $B_s\rightarrow (J/\psi,\psi(2S),\eta_c,\eta_c(2S))  K^+K^-$ decays.
Theoretical errors are attributed to the Gegenbauer moments, form factor ratios,  and hard scales, respectively.
The statistical and systematic uncertainties from data \cite{pdg2018,jhep070212017,jhep080372017} are combined in quadrature.}
\label{tab:brp}
\begin{tabular}[t]{lccc}
\hline\hline
Modes & $\mathcal{B}(R=\phi(1020))$& $\mathcal{B}(R=\phi(1680))$ & \text{P-wave}\\ \hline
$J/\psi K^+K^-$  & $5.7^{+0.2+0.9+0.2}_{-0.2-0.7-0.0}\times 10^{-4}$ & $3.6^{+0.1+0.4+0.3}_{-0.1-0.3-0.3}\times 10^{-5}$
 & $5.9_{-0.1-0.7-0.0}^{+0.1+0.9+0.3}\times 10^{-4}$  \\
Data \cite{jhep080372017} & $(5.6\pm 0.5)\times 10^{-4}$ \footnotemark[1]
&$(3.2\pm 0.4)\times 10^{-5}$ \footnotemark[1] &  $\cdots$\\
$ \eta_c K^+K^- $  & $2.4^{+0.2+0.5+0.2}_{-0.2-0.4-0.0}\times 10^{-4}$ & $1.1^{+0.1+0.2+0.0}_{-0.1-0.2-0.0}\times 10^{-5}$
& $2.4^{+0.2+0.6+0.4}_{-0.1-0.3-0.0}\times 10^{-4}$\\
Data \cite{jhep070212017} & $(2.5\pm 0.4)\times 10^{-4}$ \footnotemark[2] &  $\cdots$&  $\cdots$\\
$\psi(2S)  K^+K^- $  & $2.4^{+0.0+0.3+0.0}_{-0.1-0.4-0.0}\times 10^{-4}$ & $3.2^{+0.1+0.7+0.4}_{-0.1-0.4-0.0}\times 10^{-6}$
& $2.3^{+0.1+0.3+0.1}_{-0.1-0.4-0.0}\times 10^{-4}$\\
Data \cite{pdg2018} & $(2.6\pm 0.3)\times 10^{-4}$ \footnotemark[3] &  $\cdots$&  $\cdots$\\
$ \eta_c(2S) K^+K^- $  & $8.0^{+0.6+2.2+0.9}_{-0.6-1.8-0.5}\times 10^{-5}$ & $1.1^{+0.1+0.2+0.2}_{-0.1-0.2-0.0}\times 10^{-6}$
& $8.0^{+0.6+2.0+1.0}_{-0.7-1.8-0.5}\times 10^{-5}$ \\
\hline\hline
\end{tabular}
\footnotetext[1]{The fit fractions determined from the Dalitz plot analysis have been converted into the branching ratio measurements.}
 \footnotetext[2]{The experimental data is obtained by the product of $\mathcal{B}(B_s\rightarrow \eta_c \phi(1020))$
 and $\mathcal{B}(\phi(1020)\rightarrow K^+K^-$). }
 \footnotetext[3]{The experimental data is obtained by the product of $\mathcal{B}(B_s\rightarrow \psi(2S) \phi(1020))$
 and $\mathcal{B}(\phi(1020)\rightarrow K^+K^-$).}
\end{table}
\begin{table}
\caption{  Branching ratios  of $D$-wave resonant contributions to  the $B_s\rightarrow (J/\psi,\psi(2S),\eta_c,\eta_c(2S))  K^+K^-$ decays.
For theoretical errors, see Table. \ref{tab:brp}. The statistical and systematic uncertainties from \cite{jhep080372017} are combined in quadrature.}
\label{tab:brd}
\tiny
\begin{tabular}[t]{lccccc}
\hline\hline
Modes & $\mathcal{B}(R=f'_2(1525))$& $\mathcal{B}(R=f_2(1270))$ & $\mathcal{B}(R=f_2(1750))$ & $\mathcal{B}(R=f_2(1950))$  & \text{D-wave}\\ \hline
$J/\psi K^+K^-$   & $8.9^{+2.8+1.8+0.9}_{-2.3-1.3-0.1}\times 10^{-5}$ & $2.3^{+0.8+0.5+0.3}_{-0.5-0.4-0.1}\times 10^{-7}$
& $5.0^{+1.5+0.6+0.1}_{-1.3-0.4-0.0}\times 10^{-6}$ & $4.1^{+1.3+0.8+0.1}_{-1.0-0.7-0.0}\times 10^{-6}$
& $9.2^{+2.8+1.9+0.9}_{-2.5-1.2-0.0}\times 10^{-5}$  \\
Data \cite{jhep080372017} & $(8.5\pm 1.2)\times 10^{-5}$ & $(1.3\pm 0.3)\times 10^{-5}$
& $4.7^{+2.4}_{-2.1}\times 10^{-6}$& $3.5^{+1.7}_{-1.4}\times 10^{-6}$ &  $\cdots$\\
$\eta_c K^+K^-$  & $4.9^{+2.1+1.6+0.5}_{-1.8-1.3-0.2}\times 10^{-5}$ & $1.5^{+0.6+0.3+0.1}_{-0.5-0.3-0.1}\times 10^{-7}$
& $2.3^{+1.0+0.8+0.3}_{-0.8-0.5-0.1}\times 10^{-6}$& $2.6^{+1.1+0.4+0.1}_{-0.9-0.4-0.0}\times 10^{-6}$&
$4.9^{+2.2+1.6+0.4}_{-1.7-1.2-0.1}\times 10^{-5}$\\
$ \psi(2S)K^+K^-  $  & $1.3^{+0.4+0.2+0.0}_{-0.4-0.2-0.1}\times 10^{-5}$ & $5.9^{+1.8+1.4+0.3}_{-1.6-1.2-0.3}\times 10^{-8}$
&  $\cdots$&  $\cdots$ & $1.3^{+0.3+0.2+0.0}_{-0.4-0.2-0.1}\times 10^{-5}$\\
$\eta_c(2S)K^+K^- $  & $0.7^{+0.4+0.3+0.1}_{-0.2-0.2-0.1}\times 10^{-5}$ & $3.2^{+1.4+0.8+0.1}_{-1.1-0.7-0.1}\times 10^{-8}$
&  $\cdots$&  $\cdots$ & $0.7^{+0.3+0.3+0.1}_{-0.2-0.2-0.0}\times 10^{-5}$ \\
\hline\hline
\end{tabular}
\end{table}

The calculated branching ratios of $S$, $P$, and $D$-wave resonance contributions to
the $B_s\rightarrow (J/\psi,\psi(2S),\eta_c,\eta_c(2S))K^+K^-$
decays are collected in Tables \ref{tab:brs}, \ref{tab:brp}, and \ref{tab:brd}, respectively.
The last column of each Table are the corresponding total partial wave branching ratios.
The theoretical errors stem from the uncertainties for fitted values of Gegenbauer moments $a_i$,
the form factor ratios $r^T(R)$, and the hard scales $t$, respectively.
For Gegenbauer moments in the twist-2 DAs,
 we vary their values   within a $20\%$ range for the error estimation.
The uncertainty of the ratio $r^T(R)$ in Eqs. (\ref{eq:rt1}) and (\ref{eq:rt2}) are general assigned to be $\delta r= \pm 0.2$.
The hard scales vary from $0.75t$ to $1.25t$ to characterize the energy release in decay process.
It is necessary to stress that the second uncertainty from $r^T(R)$ is absent for the $S$-wave resonance contributions in Table \ref{tab:brs}.
The uncertainties stemming from the weight coefficients $c_R$ are not shown explicitly in these Tables,
whose effect on the branching ratios via the relation of  $\mathcal{B}\propto |c_R|^2$.
For the $S$ and $D$-waves resonance contributions,
the  twist-3 in the two-kaon DAs  are taken as the asymptotic forms for lack of better results from nonperturbative methods,
which may give significant uncertainties.
 We have checked the sensitivity of our results to the choice of the
 meson radius parameter $r$ [ see Eq.~(\ref{eq:radii})] in the BW parametrization.
The variation of its value from 0 $\text{GeV}^{-1}$ to 3.0 $\text{GeV}^{-1}$ results in the change of the branching ratios and polarizations
by only a few percents.
In general,  our results are more sensitive to those  hadronic parameters.

Before discussing the results of our calculations in detail,
 we wish to explain the quoted experimental values  that appear in these Tables.
The  measured branching ratio for each resonant component in the concerned decays
are calculated by multiplying its fit fraction and the total three-body decay branching ratio.
\footnote{So far, only the $B_s\rightarrow J/\psi K^+K^-$ mode is well measured.
Its weighted average  branching ratio, given by the Particle Data Group (PDG), is
 $\mathcal{B}(B_s\rightarrow J/\psi K^+K^-)=(7.9\pm 0.7)\times 10^{-4}$ \cite{pdg2018},
 where  the statistical and systematic uncertainties are combined in quadrature.}
The fit fractions of $P$ and $D$-wave resonances are taken from the most recent LHCb experiment \cite{jhep080372017},
which superseded the earlier one from \cite{prd87072004}.
However, in Ref.~\cite{jhep080372017}, the $S$-wave component is described in a model-independent pattern,
making no assumptions of any $f_0$ resonant structures.
Therefore, we use the $S$-wave $f_0(980)$ and $f_0(1370)$ fractions from \cite{prd87072004}.
Note that the $f_0(980)$ fraction is strongly parametrization dependent.
For instance, the  parameter set by $BABAR$ gives a smaller fit fraction $(4.8\pm 1.0)\%$,
while the parameter set by LHCb gives a larger value $(12.0\pm 1.8)\%$ [see Table VI of Ref. \cite{prd87072004}].
Therefore, we quote the central values in a wide range according to the two models
rather than a central value plus or minus its statistical and systematic
uncertainties for the $f_0(980)$ resonance in Table \ref{tab:brs}.
For other charmonium channels, the detailed partial wave analysis for determining various resonance fractions
are still missing due to a limited number of events.
The quasi-two-body branching ratios
can be  built from product of two two-body branching ratios when available in the narrow-width limit,
namely, $\mathcal{B}(B_s\rightarrow X R(\rightarrow K^+K^-))\approx \mathcal{B}(B_s\rightarrow X R)\times \mathcal{B}(R\rightarrow K^+K^-)$.
For example, we have used the experimental numbers $\mathcal{B}(B_s\rightarrow \eta_c\phi(1020))=(5.0\pm 0.9)\times 10^{-4}$ \cite{pdg2018,jhep070212017}
and $\mathcal{B}(\phi(1020)\rightarrow K^+K^-)=(49.2\pm 0.5)\%$ \cite{pdg2018} to obtain the experimental branching ratio for
$\mathcal{B}(B_s\rightarrow \eta_c\phi(1020)(\rightarrow K^+K^-))=(2.5\pm 0.4)\times 10^{-4}$, which is shown in Table \ref{tab:brp}.

It is clear that the predicted branching ratios of resonant  components are consistent with the data
except for the tensor resonance $f_2(1270)$.
From Table. \ref{tab:brd}, one can see that the predicted branching ratio of $B_s\rightarrow J/\psi f_2(1270)(\rightarrow K^+K^-)$
 is two order of magnitude smaller than the data.
We argue that the fit fraction of the $f_2(1270)$ component in $B_s\rightarrow J/\psi K^+K^-$ mode \cite{jhep080372017}
seems to be puzzling  
since it shows a tension with the corresponding one  in $B_s\rightarrow J/\psi \pi^+\pi^-$  \cite{190305530}.
\footnote{From discussions with Liming Zhang and Xuesong Liu, the $f_2(1270)$ fraction in $B_s\rightarrow J/\psi K^+K^-$ \cite{jhep080372017}
could be too high because  the misidentified background from $B_d\rightarrow J/\psi K^+\pi^-$ in the $f_2(1270)$ region may give some systematic uncertainties. }
For illustration we have explicitly written  the relative ratio of $\mathcal{B}(B_s\rightarrow J/\psi f_2(1270)(\rightarrow K^+K^-))$
compared to $\mathcal{B}(B_s\rightarrow J/\psi f_2(1270)(\rightarrow \pi^+\pi^-))$ in the narrow-width limit as
\begin{eqnarray}
\mathcal{R}=\frac{\mathcal{B}(B_s\rightarrow J/\psi f_2(1270)(\rightarrow K^+K^-))}{\mathcal{B}(B_s\rightarrow J/\psi f_2(1270)(\rightarrow \pi^+\pi^-))}\approx
\frac{\mathcal{B}(f_2(1270)\rightarrow K^+K^-)}{\mathcal{B}(f_2(1270)\rightarrow \pi^+\pi^-)},
\end{eqnarray}
in which  the  common  term $\mathcal{B}(B_s \rightarrow J/\psi f_2(1270))$ in the numerator and denominator cancel out.
It is  well known that the dominant decay mode of  $f_2(1270)$ is $\pi\pi$ rather than  the $K\bar K$,
we can thus infer that $\mathcal{R}$ should typically  be much less than 1. More specifically,
by using the numbers  $\mathcal{B}(f_2(1270)\rightarrow K\bar{K})=4.6\%$ and
$\mathcal{B}(f_2(1270)\rightarrow \pi\pi)=84.2\%$ from PDG  \cite{pdg2018}, we further get $\mathcal{R}=0.04$.
Conversely, from the Table 5 in Ref. \cite{190305530}, Table 4 in Ref. \cite{prd89092006}, and Table 3 in Ref. \cite{jhep080372017},
one can estimate  its  range from 1.9 to 4.4.
It clearly indicates that future improved measurements should take the discrepancy into account.
Assuming that the $f_2(1270)$ fraction in $B_s\rightarrow J/\psi \pi^+\pi^-$ mode  \cite{190305530} is precise enough,
combining with  above ratio $\mathcal{R}=0.04$,
we can infer $\mathcal{B}_{\text{exp}}(B_s\rightarrow J/\psi f_2(1270)(\rightarrow K^+K^-))=2.7\times 10^{-7}$.
 One can see from Table  \ref{tab:brd} that the predicted branching ratio
$\mathcal{B}(B_s\rightarrow J/\psi f_2(1270)(\rightarrow K^+K^-))\sim 2.3\times 10^{-7}$
is in agreement with the experiment.

In order to verify the validity of our numerical results, we perform a set of cross-checks.
\begin{itemize}
\item[(I)]
Using our values from Table. \ref{tab:brs}, we expect that
\begin{eqnarray}\label{eq:rr}
\frac{\mathcal{B}(B_s\rightarrow J/\psi f_0(980)(\rightarrow K^+K^-))}{\mathcal{B}(B_s\rightarrow J/\psi f_0(980)(\rightarrow \pi^+\pi^-))}
\approx \frac{\mathcal{B}( f_0(980)\rightarrow K^+K^-)}{\mathcal{B}(f_0(980)\rightarrow \pi^+\pi^-)}=0.37^{+0.23}_{-0.13},
\end{eqnarray}
where the value of $\mathcal{B}(B_s\rightarrow J/\psi f_0(980)(\rightarrow \pi^+\pi^-))=1.15^{+0.52}_{-0.41}\times 10^{-4}$
is read from the previous PQCD calculations \cite{prd91094024}.
On the experimental side, $BABAR$ measures
the ratio of the partial decay width of $f_0(980)\rightarrow K^+K^-$ to $f_0(980)\rightarrow \pi^+\pi^-$
of $0.69\pm 0.32$ using $B\rightarrow KKK$ and $B\rightarrow K\pi\pi$ decays \cite{prd74032003}.
While BES  performs a partial wave analysis of $\chi_{c0}\rightarrow f_0(980)f_0(980)\rightarrow \pi^+\pi^-\pi^+\pi^-$
and $\chi_{c0}\rightarrow f_0(980)f_0(980)\rightarrow \pi^+\pi^-K^+K^-$ in $\psi(2S)\rightarrow \gamma \chi_{c0}$ decay
and  extracts the ratio as $0.25^{+0.17}_{-0.11}$ \cite{prd72092002}.
Their weighted average yields $0.35^{+0.15}_{-0.14}$.
 It can be seen that our estimate in Eq. (\ref{eq:rr}) is consistent with this experimental average value.

\item[(II)]
Combining Tables. \ref{tab:brs}, \ref{tab:brp} and the number in Eq. (\ref{eq:rr}) , we obtain the ratio:
\begin{eqnarray}
\mathcal{R}_{f_0/\phi}=\frac{\mathcal{B}(B_s\rightarrow J/\psi f_0(980)(\rightarrow \pi^+\pi^-))}
{\mathcal{B}(B_s\rightarrow J/\psi \phi(1020)(\rightarrow K^+K^-))}=0.203^{+0.126}_{-0.095},
\end{eqnarray}
comply with the latest average of Heavy Flavor Averaging Group (HFAVG)
$\mathcal{R}_{f_0/\phi}=0.207\pm 0.016$ \cite{HFAVG} from the measurements \cite{plb698115,prd85011103,plb75684,prd84052012}
\begin{eqnarray}
\mathcal{R}_{f_0/\phi}=\left\{
\begin{aligned}
&0.252^{+0.046}_{-0.032}(\text{stat})^{+0.027}_{-0.033}(\text{syst}) \quad\quad\quad  &\text{LHCb}, \nonumber\\ 
&0.275 \pm 0.041 (\text{stat})\pm 0.061 (\text{syst}) \quad\quad\quad  & \text{D0},  \nonumber\\ 
&0.140 \pm 0.008 (\text{stat})\pm 0.023 (\text{syst}) \quad\quad\quad  & \text{CMS},  \nonumber\\ 
&0.257 \pm 0.020 (\text{stat})\pm 0.014 (\text{syst}) \quad\quad\quad  &\text{CDF}.  \nonumber\\ 
\end{aligned}\right.
\end{eqnarray}
In comparison to previous theoretical estimation  $0.122^{+0.081}_{-0.058}$ obtained in \cite{prd83094027},
our value turns out to be larger. 
\item[(III)]
Evidence of the $f_0(1370)$ resonance in  $B_s\rightarrow J/\psi \pi^+\pi^-$  decay
is reported by Belle \cite{prl106121802} with a significance of $4.2\sigma$.
The corresponding product branching fraction is measured to
$\mathcal{B}(B_s\rightarrow J/\psi f_0(1370), f_0(1370)\rightarrow \pi^+\pi^-)=
3.4^{+1.4}_{-1.5}\times 10^{-5}$ \cite{prl106121802}
\footnote{The PDG also present a value of $\mathcal{B}(B_s\rightarrow J/\psi f_0(1370)(\rightarrow \pi^+\pi^-))
=4.5^{+0.7}_{-4.0}\times 10^{-5}$ measured by the LHCb Collaboration \cite{prd86052006},
which is obtained by multiplying the corresponding normalized fit fraction
 and the  branching ratio of the normalization  mode $B_s\rightarrow J/\psi \phi(1020)$.
Although its central value is consistent with former measurements from Belle,
 but suffers  from  sizeable systematic uncertainties.
We do not use its result for further calculations.}.
Combined with our prediction on the $KK$ channel in Table. \ref{tab:brs},
one can estimate  the relative branching ratios of $f_0(1370)\rightarrow K^+K^-/\pi^+\pi^-$
lie in the range ($0.2\sim 0.5$).
Since the situation of the knowledge of the  $f_0(1370)$ decaying into $KK$ or $\pi\pi$  is rather unclear,
above expected values should be investigated further in the future with more precise data.
\item[(IV)]
From Tables \ref{tab:brp} and \ref{tab:brd}, we get another interesting ratio
\begin{eqnarray}
\mathcal{R}_{f'_2/\phi}=\frac{\mathcal{B}(B_s\rightarrow J/\psi f'_2(1525))}
{\mathcal{B}(B_s\rightarrow J/\psi \phi(1020))}=0.173^{+0.070}_{-0.059},
\end{eqnarray}
in which the two  known branching ratios  $\mathcal{B}(\phi(1020)\rightarrow K^+K^-)=(49.2\pm 0.5)\%$
and $\mathcal{B}(f'_2(1525)\rightarrow K^+K^-)=\frac{1}{2}(88.7\pm 2.2)\%$ \cite{pdg2018} are used.
 Our central value  is in accordance with the previous theoretical estimation of $0.154^{+0.090}_{-0.070}$ \cite{prd95036013}.
Experimentally,  different Collaborations  reported their measurements
$\mathcal{R}_{f'_2/\phi}=0.215\pm 0.049(\text{stat})\pm 0.026(\text{syst})$ (Belle \cite{prd88114006}),
$\mathcal{R}_{f'_2/\phi}=0.264\pm 0.027(\text{stat})\pm 0.024(\text{syst})$ (LHCb \cite{prl108151801}), and
$\mathcal{R}_{f'_2/\phi}=0.19\pm 0.05(\text{stat}) \pm 0.04(\text{syst})$ (D0 \cite{prd86092011}).
It seems that theoretical predictions are generally smaller than the experimental measurements.
None the less,
including the errors,  both the theoretical predictions and experimental data can still agree with each other.
\item[(V)] Finally, we estimate the relative branching ratios between two tensor modes
\begin{eqnarray}\label{eq:f22}
\mathcal{R}_{f_2/f'_2}=\frac{\mathcal{B}(B_s\rightarrow J/\psi f_2(1270))}
{\mathcal{B}(B_s\rightarrow J/\psi f'_2(1525))}=0.050^{+0.005}_{-0.003}.
\end{eqnarray}
The current PDG values of
\begin{eqnarray}
\mathcal{B}(B_s\rightarrow J/\psi f_2(1270)(\rightarrow \pi^+\pi^-))&=&(3.14\pm 0.9)\times 10^{-6},\nonumber\\
\mathcal{B}(B_s\rightarrow J/\psi f'_2(1525))&=&(2.6\pm 0.6)\times 10^{-4},
\end{eqnarray}
are dominated by the LHCb measurement \cite{prd86052006,prd87072004}.
Combined with the experiment value $\mathcal{B}(f_2(1270)\rightarrow \pi^+\pi^-)=\frac{2}{3}(84.2^{+2.9}_{-0.9})\%$,
one obtains the  measured ratio $\mathcal{R}_{f_2/f'_2}=0.02\pm 0.01$, which is only half of our prediction in Eq. (\ref{eq:f22}).
However, the datum for $f_2(1270)$ mode has been further reviewed in Ref \cite{190305530},
the updated branching ratio is $\mathcal{B}(B_s\rightarrow J/\psi f_2(1270)(\rightarrow \pi^+\pi^-))=(6.8 \pm 1.0)\times 10^{-6}$
with statistical uncertainty only,
corresponding to  $\mathcal{R}_{f_2/f'_2}=0.05\pm 0.01$. 
It is clear that our prediction on this ratio is marginally consistent with the updated experiment.
In addition, based on the chiral unitary approach for mesons, the authors of Ref. \cite{prd90094006}
present a larger value $\mathcal{R}_{f_2/f'_2}=0.084\pm 0.046$. 
Recalling that the  theoretical errors are relatively large, so within a $1\sigma$
tolerance, one still can count them as being consistent.
\end{itemize}

As seen in Table \ref{tab:brs}, the sum of resonance contributions from  $f_0(980)$ and
$f_0(1370)$ is somewhat larger than the $S$-wave total contribution
 due to the destructive interference between the two resonances. In fact, the best
fit model from the LHCb experiment \cite{prd87072004} also shows that the destructive interference between $f_0(980)$ and $f_0(1370)$ resonances
in the $B_s\rightarrow J/\psi K^+K^-$ channel.
The interference between the two $P$-wave resonances $\phi(1020)$ and $\phi(1680)$ is rather small
due to the relatively narrow width  of the former ($\Gamma_{\phi(1020)}=$4.25 MeV).
Since the contribution of the latter is an order of magnitude smaller, 
the $P$-wave resonance contribution is almost equal to the $\phi(1020)$ one.
By the same token,
the $D$-wave resonance contribution mainly come from the $f_2'(1525)$ component,
while other tensor resonance contributions  are at least one order smaller.
The peak of the  high-mass vector resonance $\phi(1680)$
 lie almost on the upper limit of the allowed phase space for the $2S$ charmonium modes,
their rates suffer a strong suppression 
and  are smaller than that of ground state charmonium channels by almost a factor of 10.
Higher-mass $K^+K^-$ resonances like $f_2(1750)$ and $f_2(1950)$  are beyond the invariant mass spectra
for the $2S$ charmonium modes, their contributions are absent in the last two rows of  Table \ref{tab:brd}.
As stated above, 
any interference contribution between different spin-$J$ states integrates to zero.
Therefore, summing over the contributions of the various partial waves, we can obtain the total three-body decay branching ratios
 \begin{eqnarray} \label{eq:3body}
\mathcal{B}(B_s\rightarrow J/\psi K^+K^-)&=&7.2^{+1.3}_{-0.7}\times 10^{-4}, \nonumber\\
\mathcal{B}(B_s\rightarrow \eta_c K^+K^-)&=&3.2^{+0.9}_{-0.6}\times 10^{-4},\nonumber\\
\mathcal{B}(B_s\rightarrow \psi(2S) K^+K^-)&=&2.5^{+0.4}_{-0.5}\times 10^{-4},\nonumber\\
\mathcal{B}(B_s\rightarrow \eta_c(2S) K^+K^-)&=&0.9^{+0.3}_{-0.2}\times 10^{-4},
\end{eqnarray}
in which all the uncertainties have been added in quadrature.
For the channel $B_s\rightarrow J/\psi K^+K^-$,
 the obtained branching ratio is  slightly smaller than
the current  PDG average value of $(7.9\pm 0.7)\times 10^{-4}$ \cite{pdg2018}.
Moreover, keeping in mind that  we are not including the nonresonant $S$-wave contribution in our calculations.
The small gap might be offset by the nonresonant term and its interference with the resonant components.
For other modes, their branching ratios  can also reach the order of $10^{-4}$,
which is large enough to permit a measurement.


\begin{figure}[tbp]
\begin{center}
\setlength{\abovecaptionskip}{0pt}
\centerline{
\hspace{4cm}\subfigure{\epsfxsize=13 cm \epsffile{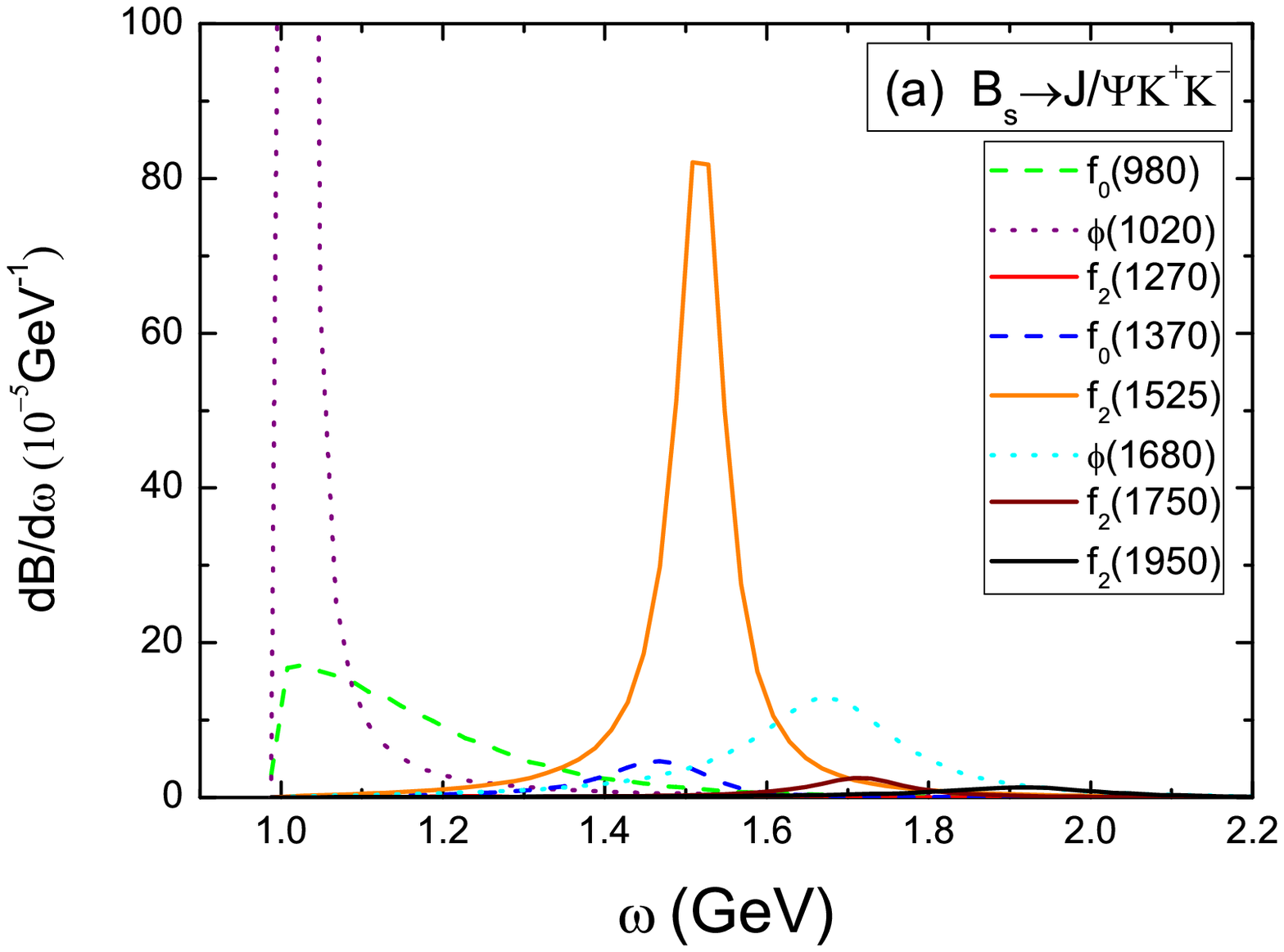} }
\hspace{-5cm}\subfigure{ \epsfxsize=13 cm \epsffile{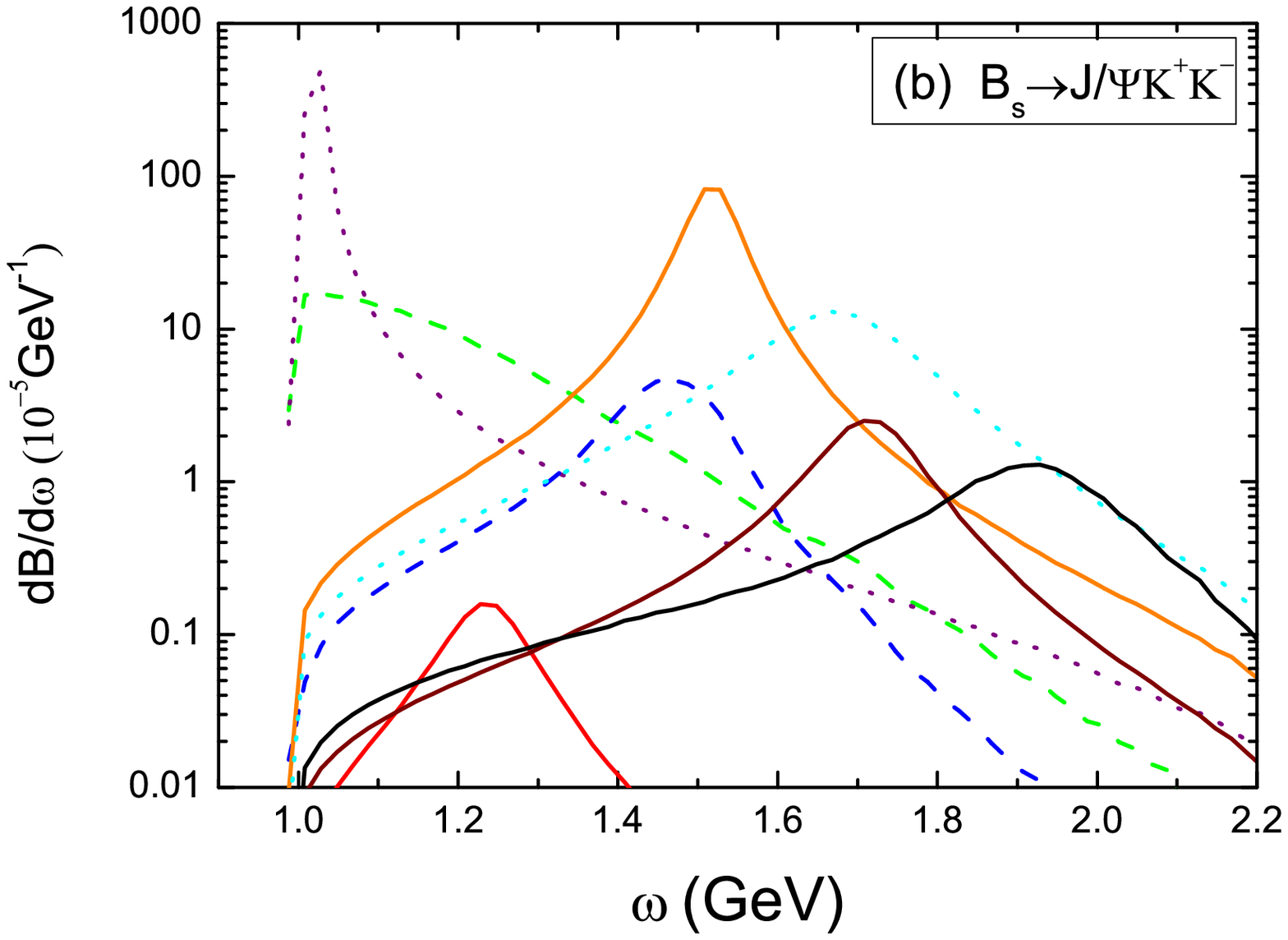}}}
\vspace{-4cm}
\centerline{
\hspace{4cm}\subfigure{\epsfxsize=13 cm \epsffile{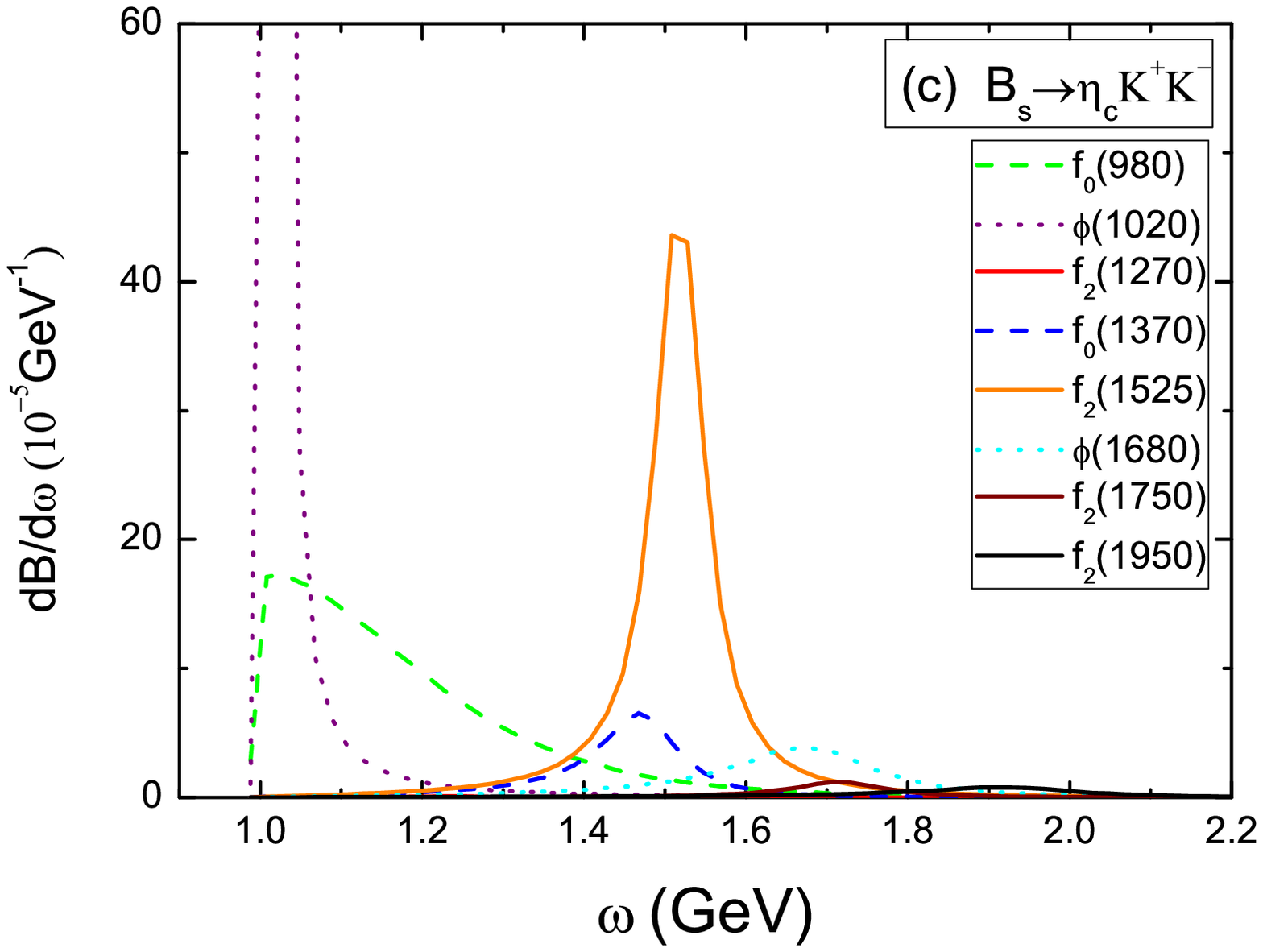} }
\hspace{-5cm}\subfigure{ \epsfxsize=13 cm \epsffile{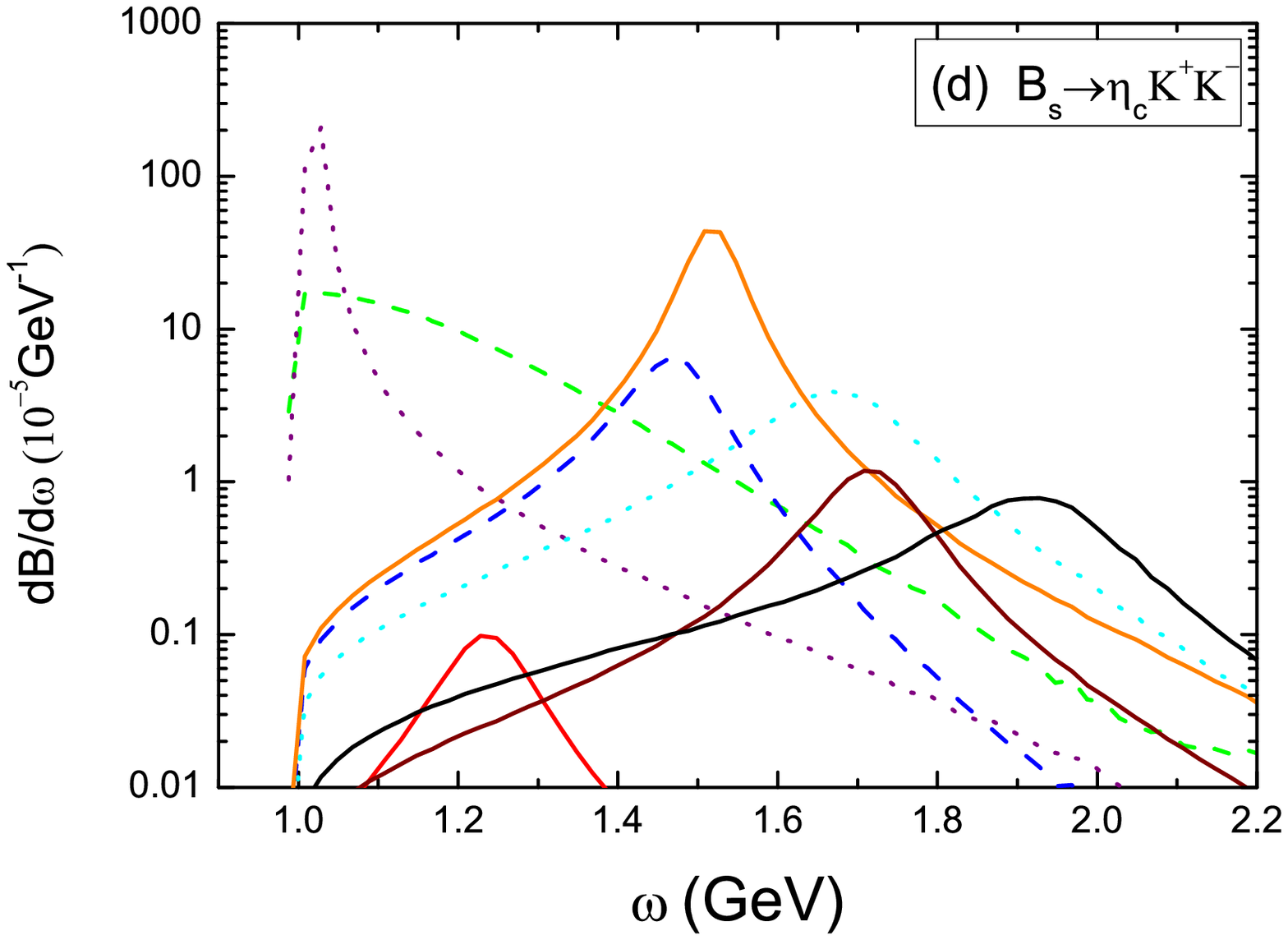}}}
\vspace{-4cm}
\centerline{
\hspace{4cm}\subfigure{\epsfxsize=13 cm \epsffile{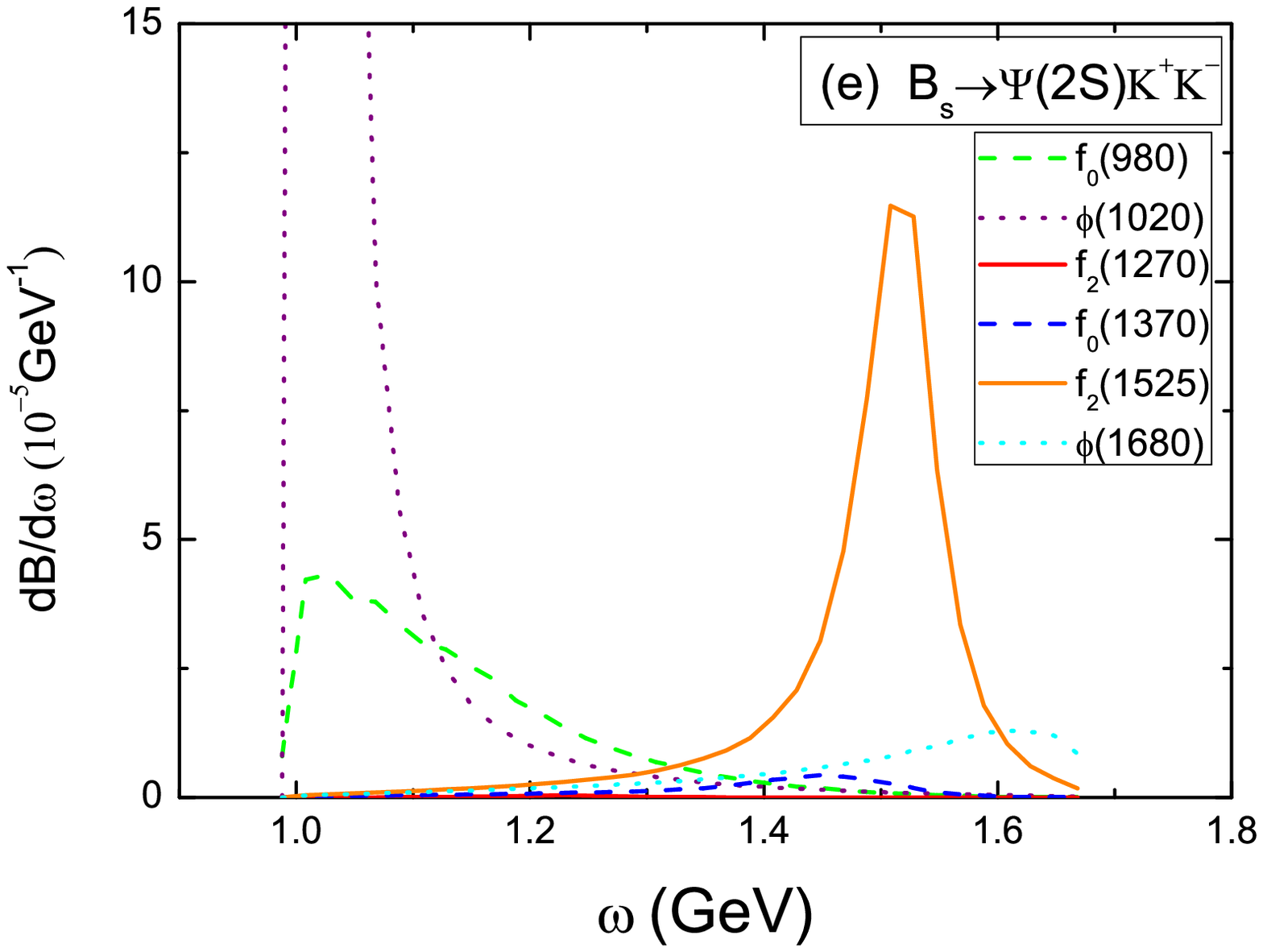} }
\hspace{-5cm}\subfigure{ \epsfxsize=13 cm \epsffile{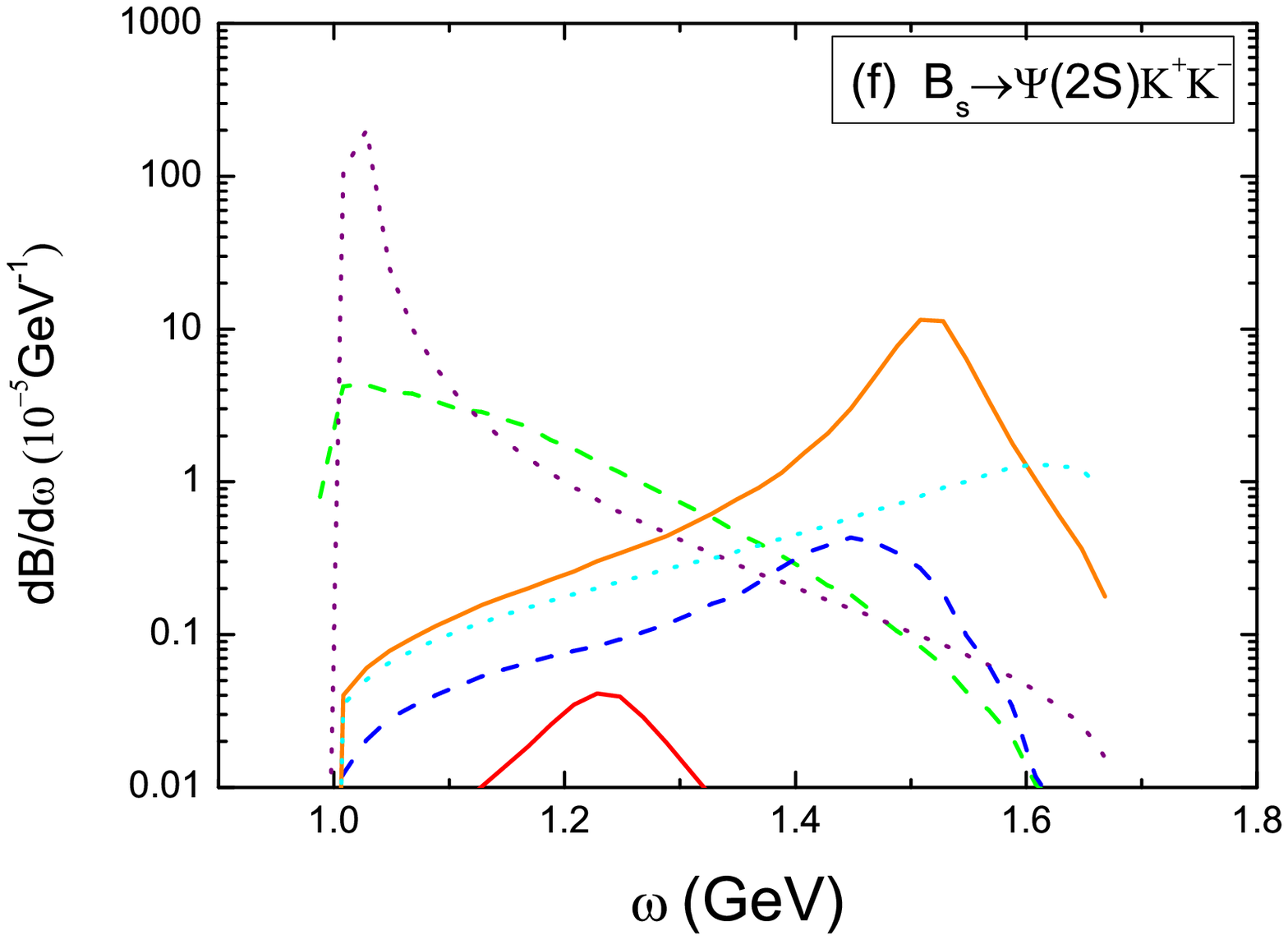}}}
\vspace{-4cm}
\centerline{
\hspace{4cm}\subfigure{\epsfxsize=13 cm \epsffile{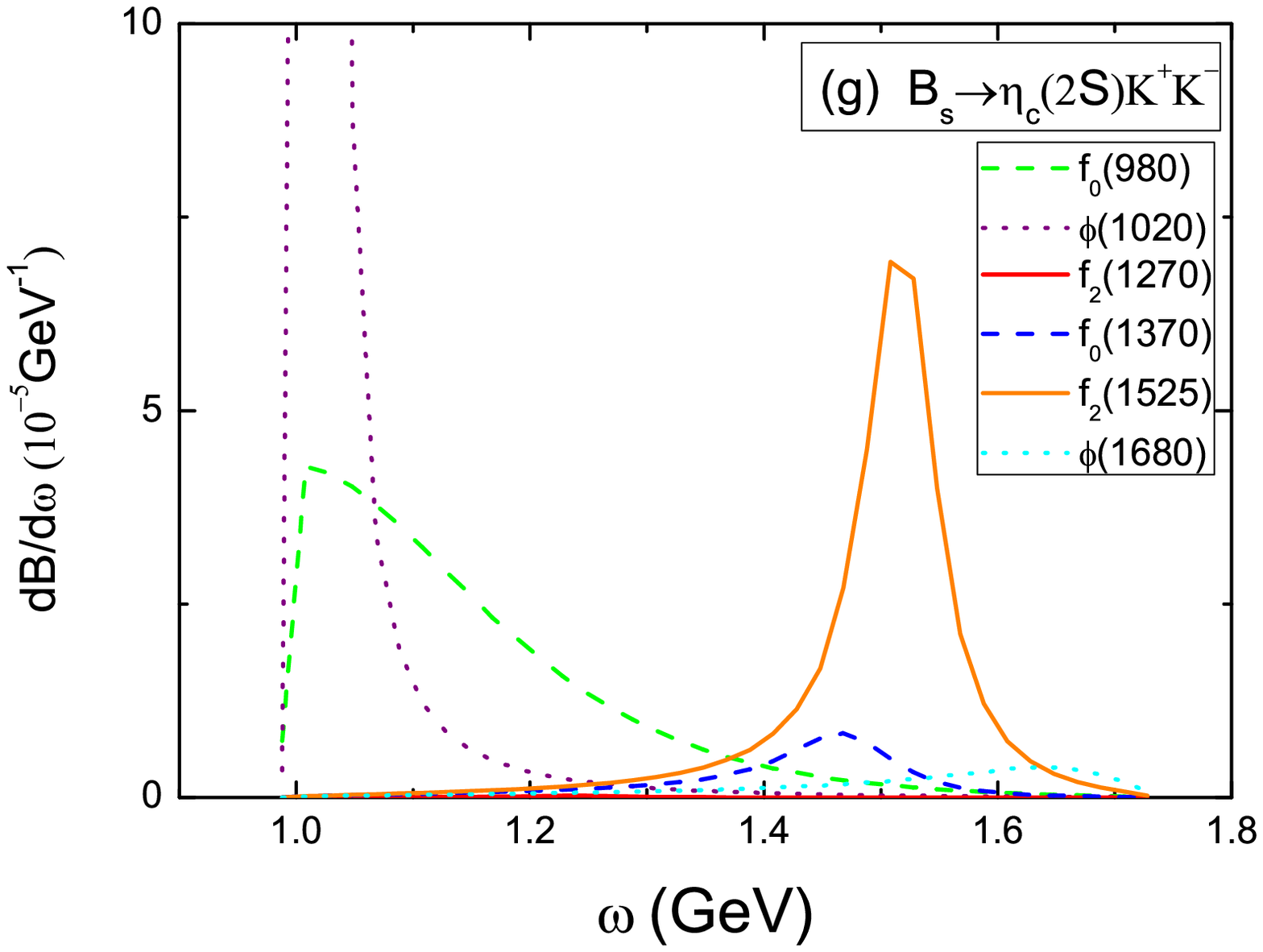} }
\hspace{-5cm}\subfigure{ \epsfxsize=13 cm \epsffile{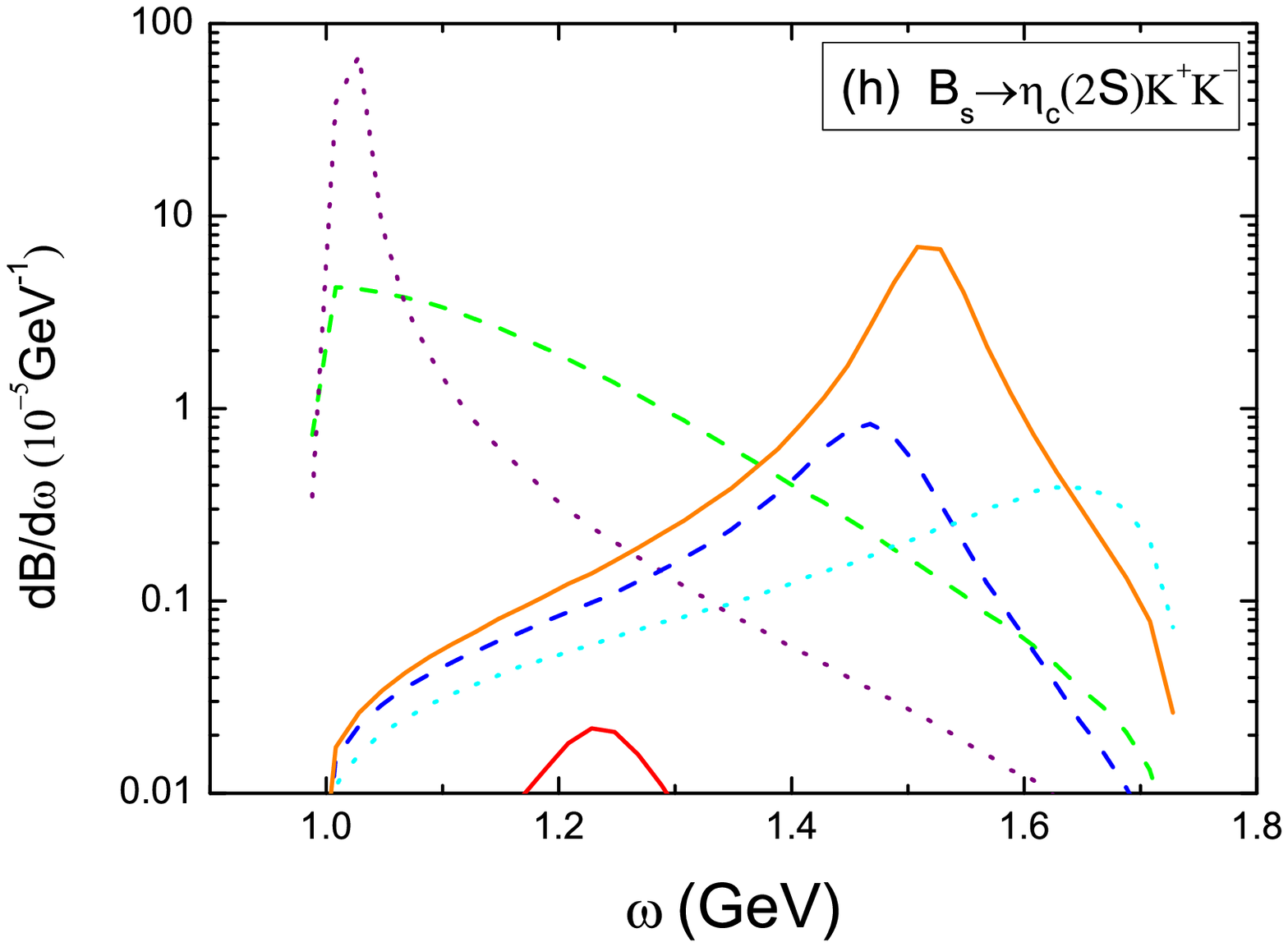}}}
\vspace{-3cm}
\caption{Various resonance contributions to the differential branching ratios of
the modes (a) $B_s\rightarrow J/\psi K^+K^-$, (c) $B_s\rightarrow \eta_c K^+K^-$,
(e) $B_s\rightarrow \psi(2S) K^+K^-$, and (g) $B_s\rightarrow \eta_c(2S) K^+K^-$ with a linear scale.
Same curves shown in (b), (d), (f), and (h) with a logarithmic scale. Components are described in the legend.}
 \label{fig:pwave}
\end{center}
\end{figure}

In the literatures, most of the theory studies concentrate on several dominant resonant components.
For example, the authors of Ref.~\cite{prd95036013} considered two dominant $\phi(1020)$ and $f'_2(1525)$
resonances in the $B_s\rightarrow J/\psi K^+K^-$ decay.
The predicted resonance contributions as well as the total three-body decay branching ration are
$(5.6\pm 0.7)\times 10^{-4}$,  $1.8^{+1.1}_{-0.8}\times 10^{-4}$, \footnote{From discussion with N\'{e}stor Quintero,
there is a typo for the $f'_2(1525)$ contribution in the Table IV of \cite{prd95036013},
its value should be $1.8$ rather than $0.8$, such that
the sum in the last column is $9.3$. } and  $9.3^{+1.3}_{-1.1}\times 10^{-4}$, respectively.
Another earlier paper \cite{prd89095026} also discuss the concerned decays in the QCD factorization approach.
The three-body branching ratio was obtained by applying Dalitz plot analysis to be
$\mathcal{B}(B_s\rightarrow J/\psi K^+K^-)=(10.3\pm 0.9)\times 10^{-4}$.
In a recent paper \cite{190602489}, the authors have performed phenomenological studies
 on the $B_s\rightarrow J/\psi f_0(980)$  decay in the two-body PQCD formalism.
  With the mixing angle between the $f_0(500)$ and  $f_0(980)$  in the quark-flavor basis  adopting as $25^{\circ}$,
  the calculated branching ratio for the two-body channel $B_s\rightarrow J/\psi f_0(980)$, was converted into quasi-two-body one as
 $\mathcal{B}(B_s\rightarrow J/\psi f_0(980)(\rightarrow K^+K^-))=4.6^{+2.6}_{-2.3}\times 10^{-5}$.
Overall, our results are comparable with these theoretical predictions within the error bars.

The differential branching ratios of the considered  decays are plotted on $\omega$ in  Fig. \ref{fig:pwave},
in which the green, purple, red, blue, orange, cyan, wine, and black lines show the $f_0(980)$, $\phi(1020)$, $f_2(1270)$, $f_0(1370)$,
$f'_2(1525)$, $\phi(1680)$,   $f_2(1750)$, and  $f_2(1950)$ resonance contributions, respectively.
To see more clearly all the resonance peaks, especially in the region of the $f_2(1270)$ resonance,
we draw them in both linear (left panels) and logarithmic (right panels) scales for each decay channel.
It is clear that an appreciable peak arising from the $\phi(1020)$ resonance, accompanied by $f_2'(1525)$.
Another three resonance peaks of $f_0(980)$, $f_0(1370)$, and $\phi(1680)$
have relatively smaller strengths than the $f_2'(1525)$ one,
but their broader widths compensate the integrated strengths over the entire phase space.
Therefore, the branching ratios of the four components 
are predicted to be of a comparable size.
Apart from above obvious signal peak, there are two visible structures at about 1750 MeV and 1950 MeV
in Fig. \ref{fig:pwave}(a) and \ref{fig:pwave}(c),
but not in  Fig. \ref{fig:pwave}(e) and \ref{fig:pwave}(g)
because the two higher mass regions are beyond the $KK$ invariant mass spectra
for the $2S$ charmonium state modes.
The contributions of the tensor $f_2(1270)$, however, can hardly be seen since
its strength is found to be compatible with zero and
its peak almost overlap with the tail of those  higher mass states like $f_2(1750)$ and $f_2(1950)$.
The obtained distribution for the most of resonance contributions to the $B_s\rightarrow J/\psi K^+K^-$ decay agrees fairly
well with the LHCb data shown in Fig.~7 of Ref.~\cite{jhep080372017},
 while other predictions could be tested by future experimental measurements.


Let us now proceed to the polarization fractions which are defined as
\begin{eqnarray}\label{pol}
f_{\sigma}=\frac{|\mathcal{A}_{\sigma}|^2}{|\mathcal{A}_0|^2
+|\mathcal{A}_{\parallel}|^2+|\mathcal{A}_{\perp}|^2},
\end{eqnarray}
with $\sigma=0,\parallel,\perp$ being the longitudinal, perpendicular,
and parallel polarizations, respectively.

\begin{table}
\caption{  Polarization fractions  for the decays $B_s\rightarrow (J/\psi, \psi(2S))\phi/f_2(\rightarrow K^+K^-)$.
For theoretical errors, see Table.~\ref{tab:brp}.
 The experimental data are taken from Ref. \cite{jhep080372017},
where  the statistical and systematic uncertainties are combined in quadrature.}
\label{tab:pol}
\begin{tabular}[t]{lccc}
\hline\hline
Modes  & $f_0(\%)$ & $f_{\parallel}(\%)$ &$f_{\perp}(\%)$\\
\hline
$B_s\rightarrow J/\psi \phi(1020)(\rightarrow K^+K^-)$  & $50.6^{+1.2+5.9+2.5}_{-1.5-4.4-1.7}$ &$24.4^{+1.1+1.8+1.0}_{-0.6-2.1-1.3}$& $24.9^{+0.6+2.7+0.9}_{-0.5-3.6-1.1}$ \\
 Data \cite{jhep080372017} & $50.9 \pm0.4 $&$23.1\pm0.5$& $26.0\pm0.6$ \\
$B_s\rightarrow J/\psi \phi(1680)(\rightarrow K^+K^-)$  & $49.1^{+1.5+3.8+0.6}_{-1.1-3.6-0.9}$ &$20.1^{+0.2+0.3+0.7}_{-0.5-0.2-0.3}$& $30.8^{+0.9+3.3+0.1}_{-1.1-3.6-0.3}$ \\
Data \cite{jhep080372017} & $44.0\pm3.9 $&$32.7\pm3.6$& $23.3\pm3.6$ \\
$B_s\rightarrow \psi(2S) \phi(1020)(\rightarrow K^+K^-)$  & $44.1^{+1.3+3.8+0.6}_{-2.4-5.0-1.8}$
&$23.2^{+1.2+1.8+1.1}_{-0.8-1.3-0.4}$&$32.7^{+1.1+3.1+0.7}_{-0.5-2.4-0.2}$ \\
Data \cite{plb762253} &$42.2\pm1.4$ &  $\cdots$ & $26.4 \pm 2.4$\\
$B_s\rightarrow \psi(2S) \phi(1680)(\rightarrow K^+K^-)$  & $45.3^{+0.3+3.5+0.2}_{-0.4-3.3-1.3}$
&$15.1^{+0.2+0.3+0.9}_{-0.1-0.2-0.3}$& $39.6^{+0.1+3.5+0.4}_{-0.1-3.8-0.0}$ \\ \hline
$B_s\rightarrow J/\psi f_2'(1525)(\rightarrow K^+K^-)$ & $51.1^{+14.1+4.3+1.6}_{-14.2-3.6-0.3}$
&$26.2^{+7.6+1.9+0.0}_{-7.6-2.1-0.4}$& $22.7^{+6.7+1.7+0.6}_{-6.5-2.2-1.4}$ \\
Data \cite{jhep080372017} & $46.8\pm1.9$&$33.8\pm2.3$& $19.4\pm2.8$ \\
$B_s\rightarrow J/\psi f_2(1270)(\rightarrow K^+K^-)$&$42.9^{+12.5+3.0+2.3}_{-14.2-2.7-2.4}$
&$29.5^{+7.4+1.3+0.6}_{-6.4-1.5-0.6}$& $27.6^{+6.7+1.4+1.8}_{-6.1-1.6-1.7}$ \\
Data \cite{jhep080372017} & $76.9\pm5.5$&$6.0\pm4.2$& $17.1\pm5.0$ \\
$B_s\rightarrow J/\psi f_2(1750)(\rightarrow K^+K^-)$ & $53.7^{+13.9+1.8+2.1}_{-14.4-3.3-1.7}$
 &$25.3^{+7.8+1.6+0.8}_{-7.6-0.7-0.3}$& $21.0^{+6.5+1.7+0.9}_{-6.3-1.0-1.9}$ \\
Data  \cite{jhep080372017}& $58.2\pm13.9$&$31.7\pm12.4$& $10.1^{+16.8}_{-6.1}$ \\
$B_s\rightarrow J/\psi f_2(1950)(\rightarrow K^+K^-)$ & $30.2^{+13.1+3.1+1.8}_{-11.1-1.2-0.0}$
&$36.9^{+5.8+0.6+0.6}_{-6.9-1.5-1.4}$& $32.9^{+5.2+0.6+0.8}_{-6.2-1.6-2.5}$ \\
Data  \cite{jhep080372017}& $2.2^{+6.7}_{-1.5}$&$38.3\pm13.8$& $59.5\pm14.2$ \\
$B_s\rightarrow \psi(2S) f_2'(1525)(\rightarrow K^+K^-)$ & $41.9^{+14.2+1.8+1.0}_{-13.5-2.7-1.9}$
&$34.7^{+8.0+2.8+0.6}_{-8.5-2.2-0.0}$& $23.4^{+5.5+0.5+1.5}_{-5.7-0.2-1.5}$ \\
$B_s\rightarrow \psi(2S) f_2(1270)(\rightarrow K^+K^-)$ &$36.4^{+13.9+3.2+0.9}_{-12.4-2.1-1.0}$
 &$37.0^{+7.3+1.1+0.8}_{-8.1-1.7-0.6}$& $26.5^{+5.2+1.0+1.7}_{-5.8-1.4-1.6}$ \\
\hline\hline
\end{tabular}
\end{table}
The PQCD results for the  polarization fractions
together with the LHCb data, are listed in Table \ref{tab:pol}.
The sources of the errors in the numerical estimates have the same origin as
in the discussion of the branching ratios in Table. \ref{tab:brp}.
For most modes, the transverse polarization fraction $f_T=f_{\parallel}+f_{\perp}$ and the longitudinal one are roughly equal.
Nevertheless, for the  $f_2(1950)$ mode, the longitudinal polarization fraction is suppressed to $30\%$
owing to a larger  $r^T(f_2(1950))$ in Eq.~(\ref{eq:rt2}) enhances its transverse polarization contribution.
Even so, the longitudinal polarization fraction is still larger than the experimental value.
Of course, taking into account both the theoretical and experimental errors, the deviation is less than $3\sigma$.

For the $P$-wave resonant channels, the parallel polarization fractions are slightly smaller
than the corresponding perpendicular one in our calculations,
while the  LHCb's data show  an opposite behavior  for the $\phi (1680)$ mode.
As pointed out in Ref.~\cite{prd98113003}, the relative importance of
the parallel and perpendicular polarization amplitudes in the $\rho$ channels
are sensitive to the two Gegenbauer moments $a_2^{a}$ and $a_2^{v}$.
The similar situation also exist  in this work.
Strictly speaking, the Gegenbauer moments in two-hadron DAs are not constants,
but depend on the dihadron invariant mass $\omega$.
However, the explicit behaviors of those Gegenbauer moments with the $\omega$ are still unknown and
the available data are not yet sufficiently precise to control their dependence.
Here, we do not consider the $\omega$ dependence and  assume
the Gegenbauer moments for the resonances with same spin are universal.
That is to say  it is unlikely to accommodate the measured $B_s\rightarrow J/\psi \phi(1020), J/\psi \phi(1680)$
parallel and perpendicular polarization simultaneously with the same set of Gegenbauer moments in PQCD.
A further theoretical study of the $\omega$ dependence of the Gegenbauer moments will clarify this issue.

For the $D$-wave mode $B_s\rightarrow J/\psi f_2(1270)$,
compared with the data from the LHCb, our predicted  longitudinal polarization is
smaller while the two transverse ones are larger [see Table \ref{tab:pol}].
As stressed before, the $f_2(1270)$ fit fraction in the $KK$ mode is unexpected,
so its polarizations may have a similar situation.
In fact, the best fit model from LHCb \cite{prd89092006,190305530} on the $B_s\rightarrow J/\psi \pi^+\pi^-$ decay showing
the longitudinal polarization for the $f_2(1270)$ component is obviously smaller than the transverse ones.
As it is hard to understand why the polarization patterns of $f_2(1270)$ resonance decaying into $\pi\pi$ and $KK$ pairs are so different,
a refined measurement of the $f_2(1270)$ contribution to the  $J/\psi K^+K^-$ mode
is urgently needed in order to clarify such issue.

So far, there are several literatures \cite{prd89094010,prd95036013,epjc77610} on the calculation of polarization fractions,
focusing more on the $\phi(1020)$ and $f'_2(1525)$ channels.
 We found numerically that
 \begin{eqnarray}
J/\psi\phi(1020)&:&\quad f_0=(50.7\pm 3.6)\%, \quad f_{\parallel}=(29.8^{+2.3}_{-2.0})\%
, \quad f_{\perp}=(19.4^{+1.7}_{-1.5})\% 
\nonumber\\
\psi(2S)\phi(1020)&:&\quad f_0=(48^{+5}_{-6})\%, \quad f_{\parallel}=(29^{+2}_{-3})\%
, \quad f_{\perp}=(24\pm 4)\% 
\nonumber\\
J/\psi f'_2(1525)&:& \quad f_0=(53.3\pm 18.0)\%, \quad f_{\parallel}=(30.8\pm 12.0)\%
, \quad f_{\perp}=(15.8\pm 0.60)\%. 
\end{eqnarray}
 It is clear from Table~\ref{tab:pol} that our calculations are comparable with theirs within errors.
Since the higher mass intermediate states in the concerned decays
are still received less attention in both theory and experiment,
we wait for future comparison.

\section{ conclusion}\label{sec:sum}
In this paper we carry out an systematic analysis of the $B_s$ meson decaying into charmonia and $K^+K^-$ pair by using the PQCD approach.
This type of process is expected to
receive  dominant  contributions from intermediate resonances,
such as the vector $\phi(1020)$, tensor $f'_2(1525)$, and scalar $f_0(980)$,
 thus  can be considered as quasi-two-body decays.
In addition to the  three prominent components mentioned above,
some significant excitations in the entire $K^+K^-$ mass spectrum,
which have been  well established in the $B_s\rightarrow J/\psi K^+K^-$ decay, are also included.
These resonances  fall into three partial waves according to their spin, namely, $S$, $P$, and $D$-wave states.
Each partial wave contribution is parametrized into the corresponding timelike form factor involved in the two-kaon DAs,
which can be described by the coherent sum over resonances sharing the same spin.
The $f_0(980)$ component is described  by a Flatt\'{e} line shape,
while other resonances are modeled by the  Breit-Wigner function.

After determining the hadronic parameters involved in the two-kaon DAs by fitting our formalism to the   available data,
we have calculated each resonance contribution  in the processes under consideration.
It is found that the largest component  is the $\phi(1020)$,
followed by $f'_2(1525)$, with others being almost an order of magnitude smaller.
The resultant invariant mass distributions for most resonances in the $B_s\rightarrow J/\psi K^+K^-$
decay show a similar qualitative behavior as the LHCb experiment.
Since the interference contributions between any two different spin resonances are zero,
summing over various partial wave contributions, we can estimate the total three-body decay branching ratios.
The obtained  branching ratio of the $B_s\rightarrow J/\psi K^+K^-$ decay is in accordance with
available experimental data and numbers from other approaches.
The modes involving  2S charmonium have sizable three-body branching ratios, of order $10^{-4}$,
which seem to be in the reach of future experiments.
As a cross-check, we have  discussed some interesting relative branching ratios and
compared with available experimental data and other theoretical predictions. 

Three polarization contributions were also investigated in detail for the vector-vector and vector-tensor modes.
For most of channels, the transverse polarization is found to be of the same size as the longitudinal one
and the  parallel and perpendicular polarizations are also roughly equal,
while for some higher resonance modes, the polarization patterns can be different.
The obtained results can be confronted to the experimental data in the future.

Finally, we emphasize that  further experimental investigations on
the  $f_2(1270)$ component in the $B_s \rightarrow J/\psi K^+K^-$ decay based on much larger data samples are urgently necessary.

\begin{acknowledgments}
We acknowledge Hsiang-nan Li for helpful discussions
and Liming Zhang for enlightening discussions concerning the experiments.
This work was supported in part by the National Natural Science Foundation of China
under Grants No.11605060 and No.11547020, in part by Natural Science Foundation of Hebei Province
under Grant no. A2019209449, and in part  by the Program for the Top Young
Innovative Talents of Higher Learning Institutions of Hebei Educational
Committee under Grant No. BJ2016041. 	
Ya Li is also supported by the Natural Science Foundation of Jiangsu under Grant No. BK20190508.
\end{acknowledgments}

\begin{appendix}
\section{the decay Amplitudes for $B_s\rightarrow \eta_c R(\rightarrow K^+K^-)$}\label{sec:apps}
The decay amplitude can be conventionally written as
\begin{eqnarray}
\mathcal{A}&=&\frac{G_F}{\sqrt{2}}\Big\{V^*_{cb}V_{cs}
\Big [(C_1+\frac{1}{3}C_2)\mathcal{F}^{LL}+C_2\mathcal{M}^{LL} \Big]
-V^*_{tb}V_{ts}\Big [(C_3+\frac{1}{3}C_4+C_9+\frac{1}{3}C_{10})\mathcal{F}^{LL}+\nonumber\\
&&(C_5+\frac{1}{3}C_6+C_7+\frac{1}{3}C_{8})\mathcal{F}^{LR}
+(C_4+C_{10})\mathcal{M}^{LL}+(C_6+C_8)\mathcal{M}^{SP}\Big ]\Big\},
\end{eqnarray}
with the Cabibbo-Kobayashi-Maskawa (CKM) matrix elements $V_{ij}$ and the Fermi coupling constant $G_F$.
$\mathcal{F}(\mathcal{M})$ describes the contributions from the factorizable (nonfactorizable ) diagrams in Fig \ref{fig:femy}.
The superscript $LL$, $LR$, and $SP$  refer to the contributions from $(V-A)\otimes(V-A)$, $(V-A)\otimes(V+A)$,
and $(S-P)\otimes(S+P)$ operators, respectively.
Performing the standard PQCD calculations, one gets the following expressions:
\begin{eqnarray}
\mathcal{F}^{LL}&=&8\pi C_F f_{\eta_c} M^4 \int_0^1dx_B dz \int_0^{\infty} b_B db_Bbdb\phi_B(x_B,b_B)\nonumber\\&&
\{[\phi^0(r^2 (-2 \eta  (z+1)+2 z+1)+(\eta -1) (z+1))+
\sqrt{\eta(1-r^2)}(\phi^s(\eta+r^2(2(\eta-1)z-1)-2(\eta-1)z-1)\nonumber\\&&
+\phi^t(\eta+r^2(2(\eta-1)z+1)-2(\eta-1)z-1))]E_e(t_a)h_a(x_B,z,b_B,b)+\nonumber\\&&
[2\phi^s(\sqrt{\eta(1-r^2)}(\eta +r^2(-2\eta+x_B+1)-1))
+\phi^0(\eta +\eta^2( r^2-1)-r^2 x_B)]E_e(t_b)h_b(x_B,z,b_B,b)\},
\end{eqnarray}
\begin{eqnarray}
\mathcal{M}^{LL}&=&-16\sqrt{\frac{2}{3}}\pi C_F  M^4 \int_0^1dx_B dz dx_3\int_0^{\infty} b_B db_Bb_3db_3\phi_B(x_B,b_B)\psi^v(x_3,b_3)
\nonumber\\&&[\phi^0(\eta+r^2-1)+2\phi^t\sqrt{\eta(1-r^2)}]
[r^2(x_B+(\eta-1)z)-\eta z+z]E_n(t_d)h_d(x_B,z,x_3,b_B,b_3),
\end{eqnarray}
\begin{eqnarray}
\mathcal{F}^{LR}=-\mathcal{F}^{LL}, \quad \mathcal{M}^{SP}=\mathcal{M}^{LL},
\end{eqnarray}
with color factor $C_F=4/3$. $f_{\eta_c}$ is the decay constant of the $\eta_c$ meson.
The expressions for the evolution functions $E$, the hard kernels $h$, and the hard scales $t_{a,b,c,d}$
 are referred to the Appendix of Ref. \cite{prd91094024}.
 The forms of $\psi^v$ are adopted as our previous works \cite{prd90114030,epjc75293}.
It should be stressed that above factorization formulas are the same for the scalar and vector resonances involved modes
 except for  their different two-kaon DAs.
The decay amplitude of the tensor modes should be multiplied by an extra factor $\sqrt{2/3}$,
which  derives from  the different definitions of the polarization vector as aforementioned.
In addition, we also consider the vertex corrections to the factorizable diagrams in Fig. \ref{fig:femy},
whose effects are absorbed into the modified Wilson coefficients as usual \cite{bbns1,bbns2,bbns3}.
 For the  calculation of vertex corrections,  one refer to \cite{prd65094023,prd63074011} for details.
The characteristic scale $\Lambda^5_{\text{QCD}}=0.225$ GeV at next-to-leading order was used in this work.
\end{appendix}

\end{document}